\documentclass{PoS}
\usepackage{amsmath}
\usepackage{booktabs}
\usepackage{color}
\usepackage{subcaption}
\graphicspath{{figs/}}
\renewcommand{\O}{\mathcal{O}}
\newcommand{\Vcb}{V_{cb}}
\newcommand{\bra}[1]{\langle #1 |}
\newcommand{\ket}[1]{| #1 \rangle}
\definecolor{dkgreen}{rgb}{0.0,0.4,0.0}
\newcommand{\green}[1]{{\color{dkgreen} #1}}
\def\[{\left[}
\def\]{\right]}
\def\({\left(}
\def\){\right)}

\title{Lattice $B \to D^{(*)}$ form factors, $R(D^{(*)})$, and $|V_{cb}|$}

\ShortTitle{$B \to D^{(*)}$ form factors}

\author{\speaker{Andrew Lytle} \\
        INFN, Sezione di Roma Tor Vergata, Via della Ricerca Scientifica 1,
        00133 Roma RM, Italy\\
        E-mail: \email{andrew.lytle@roma2.infn.it}}

\abstract{
I discuss recent 
progress in lattice calculations of $B \to D^{(*)} \ell \nu$ 
form factors, important for the precision
determination of $|V_{cb}|$ in the Standard Model (SM), 
and for testing SM expectations of lepton flavor universality
in observables $R(D^{(*)})$. 
I also discuss progress in calculations of the related 
$b \to c$ semileptonic decays 
$B_s \to D_s^{(*)} \ell \nu$ and $B_c \to J/\psi \, \ell \nu$
now experimentally accessible at the LHC.
}

\FullConference{
    The 37th Annual International Symposium 
    on Lattice Field Theory - LATTICE2019\\
	16-22 June, 2019\\
	Wuhan, China.}

\begin{document}
\section{Introduction} \label{Introduction}
\begin{figure}[t]
\includegraphics[width=0.48\textwidth]{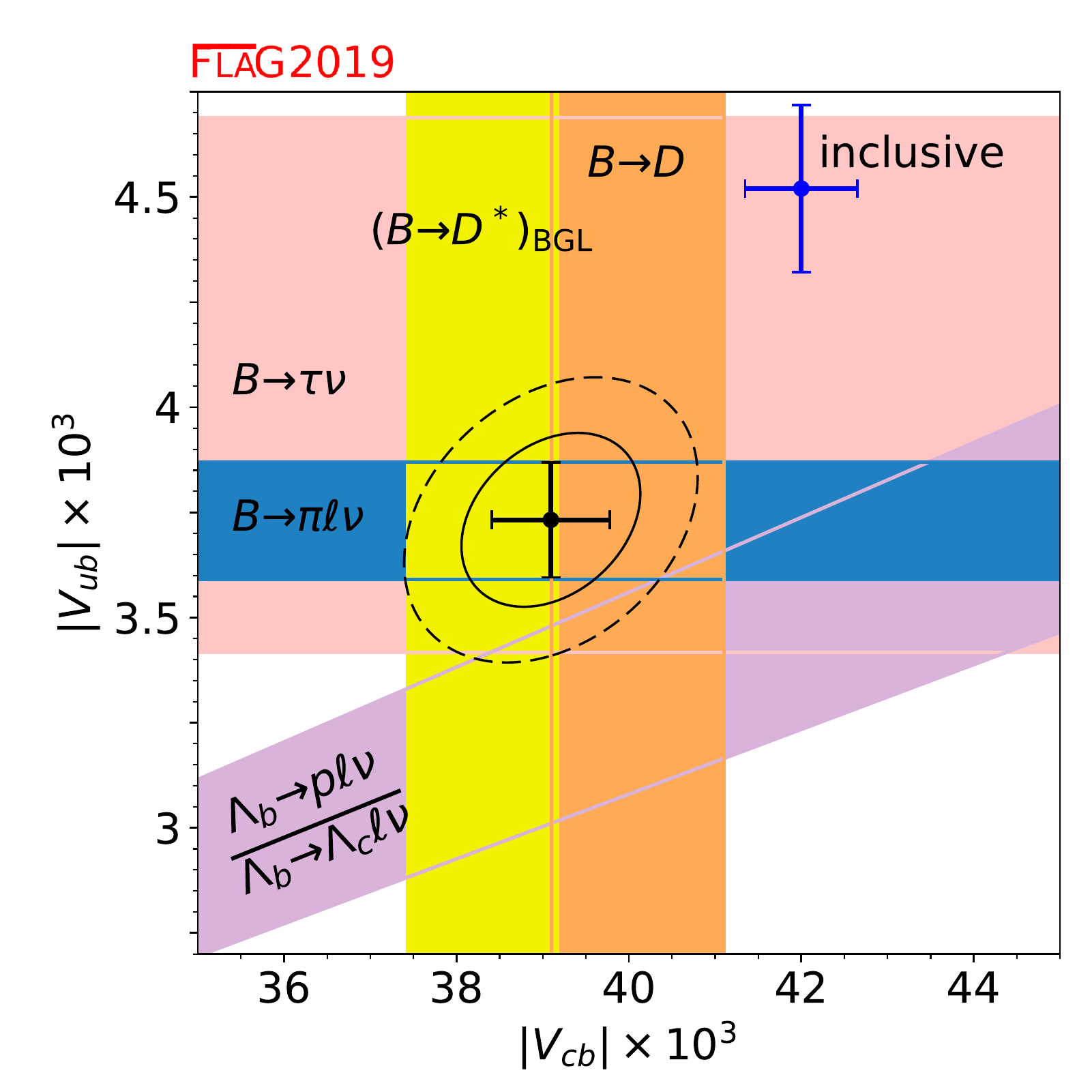} 
\includegraphics[width=0.48\textwidth]{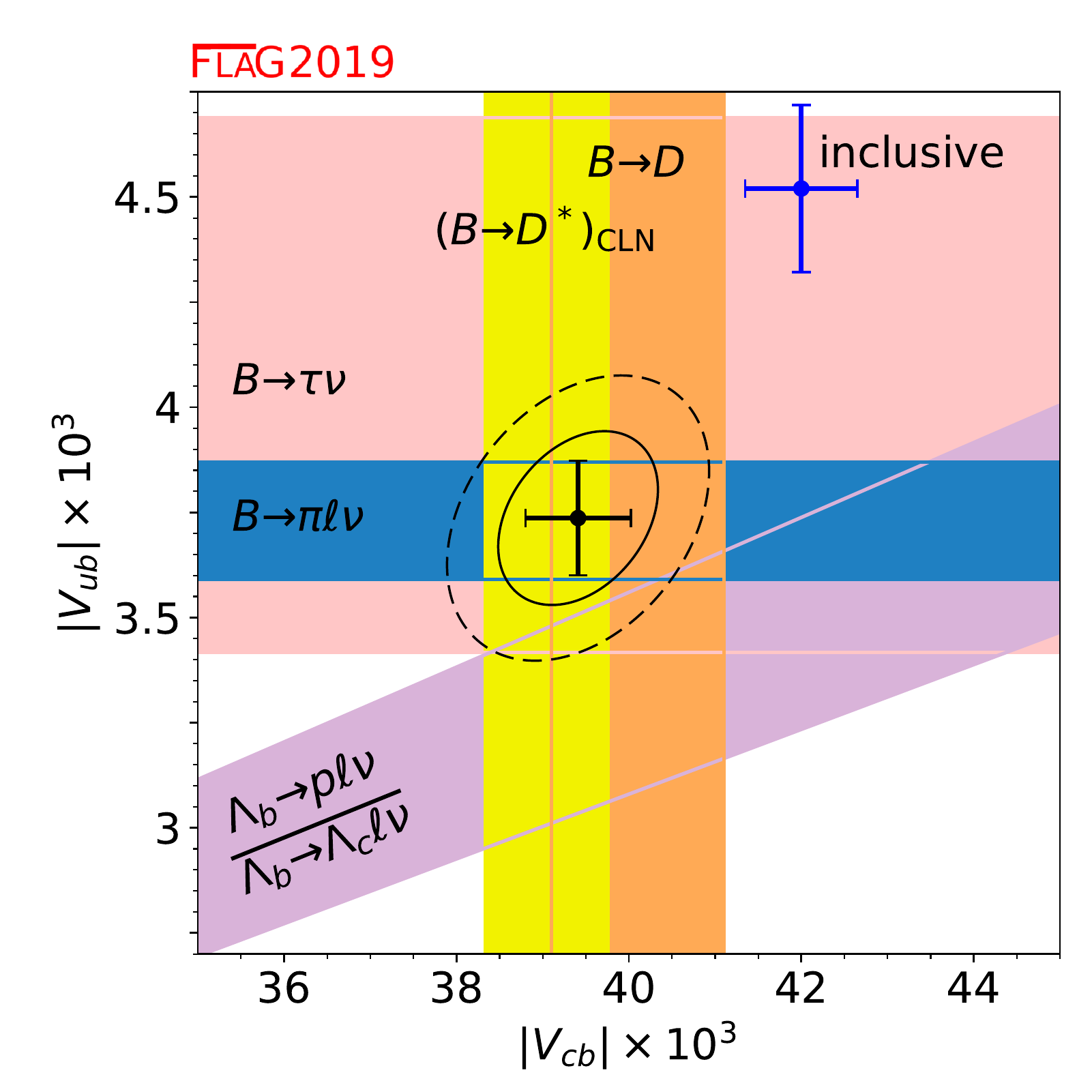}
\caption{Figures from the 2019 FLAG review \cite{Aoki:2019cca} showing
the present status of $|\Vcb|$ using various exclusive
channels compared with inclusive determinations. The vertical yellow
band is the result from $B \to D^* l \nu$ decays using either the
BGL (left) or CLN (right) parameterisations to fit the data.
\label{fig:flag-1}}
\end{figure}

The $B$-meson semileptonic decays $B \to D^{(*)} l \nu$
provide a precise way to determine the CKM matrix element $|V_{cb}|$.
In addition to experimental data, 
these determinations require the precision calculation of nonperturbative
form factors using lattice QCD. There is a long-standing discrepancy between
the values obtained from 
these exclusive determinations $|V_{cb}|^{\text{excl}}$, and those obtained
from inclusive determinations $|V_{cb}|^{\text{incl}}$ -- this is known 
as the $\Vcb$ puzzle~\cite{Gambino:2019sif}. A recent
comparison of inclusive and exclusive determinations of $|\Vcb|$ 
from the Flavor Lattice Averaging Group (FLAG) is presented 
in Fig.~\ref{fig:flag-1}.
\footnote{A review talk
summarising the status of the full CKM matrix
was given at this conference by Steve Gottlieb~\cite{Gottlieb:2020zsa}.}

\begin{figure}[t]
\includegraphics[width=0.47\textwidth]{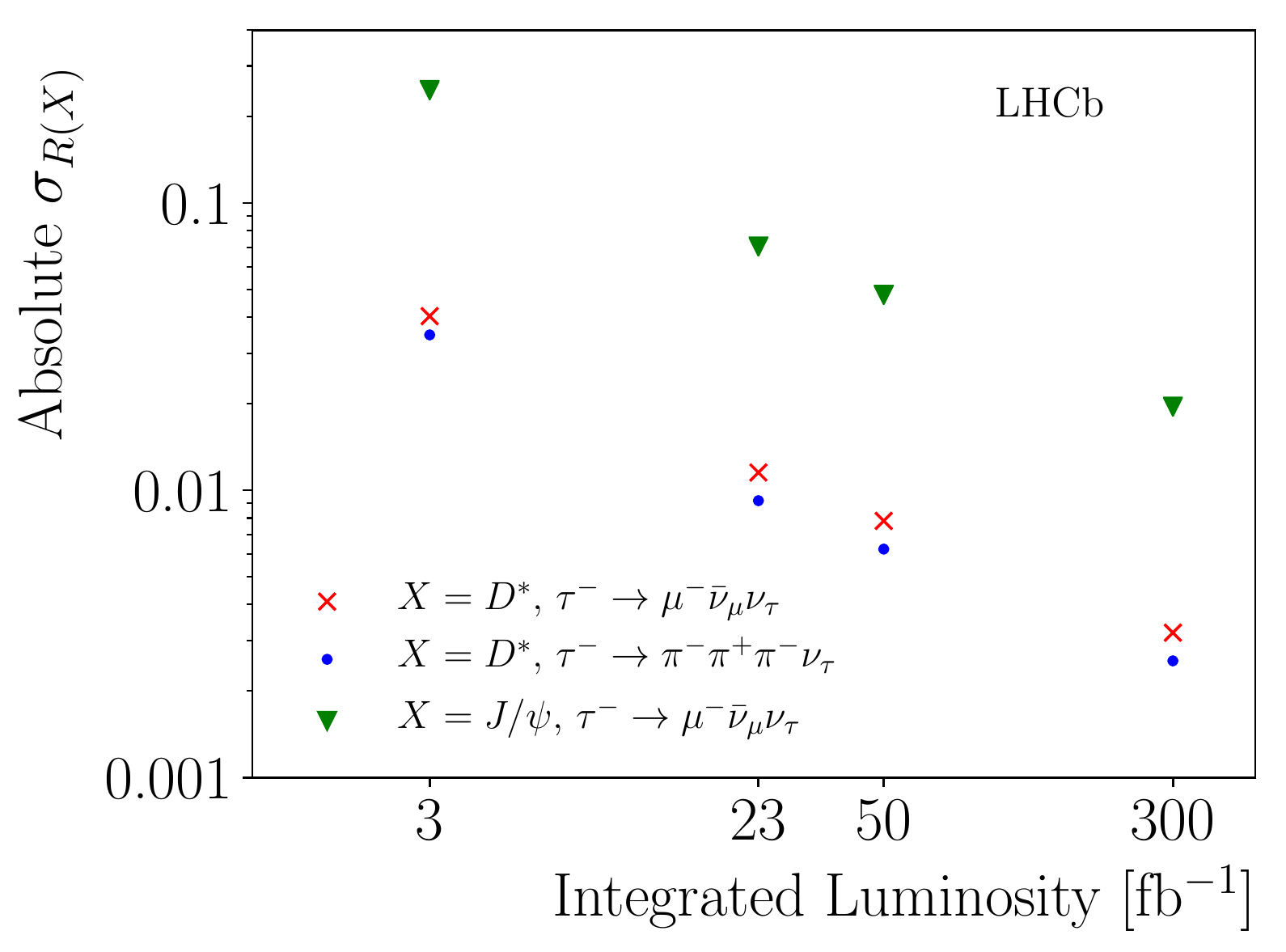}
\includegraphics[width=0.53\textwidth]{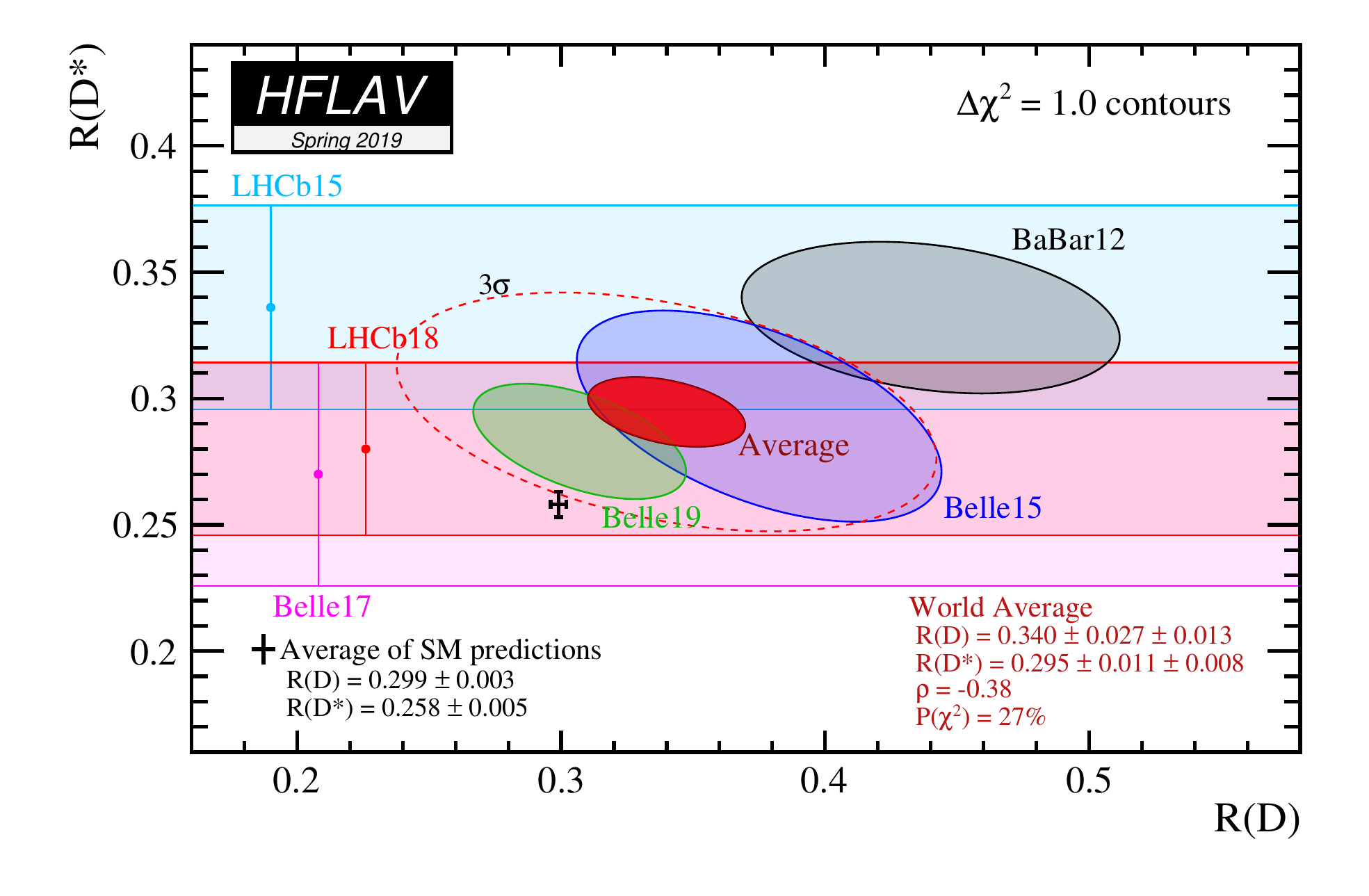}
\caption{(Left) Projections for
the expected uncertainty achievable at LHCb for ratios
$R(D^{(*)})$ and $R(J/\psi)$, reproduced from~\cite{Bediaga:2018lhg}.
(Right) Summmary of experimental status of $R(D^{(*)})$
measurements compiled by HFLAV~\cite{Amhis:2019ckw}.
\label{fig:RX_projection_stat_syst}}
\end{figure}

There are also long-standing few-sigma discrepancies with Standard Model (SM)
predictions in the measured `$R$-ratios' for these decays.
The $R$-ratio for a semileptonic
decay is defined as the branching fraction for that decay into the
tau channel divided by that for the muon or electron,
\begin{equation}
R(D^{(*)}) = \frac{\mathcal{B}(B \to D^{(*)} \tau \nu_{\tau})}
{\mathcal{B}(B \to D^{(*)} l \nu_l)} \, \quad l=\mu,\,e \,.
\end{equation}
These ratios are independent of $|\Vcb|$, but depend on the 
nonperturbative form factors over the entire kinematic range.
Recently a measurement
from LHCb found that $R(B_c \to J/\psi)$ also differs significantly from
its SM expectation~\cite{Aaij:2017tyk}. A recent synopsis of the situation for
$R(D^{(*)})$ from the Heavy Flavor Averaging Group (HFLAV) is reproduced
in Fig.~\ref{fig:RX_projection_stat_syst}.

At the same time, 
a great deal of new experimental information is expected to become available
in the near future, both at Belle II~\cite{Kou:2018nap} 
and from the LHC~\cite{Bediaga:2018lhg}. This will
lead to increasingly precise experimental information for $B \to D^{(*)}$,
and information about newly accessible decays at LHC in channels
$B_s \to D_s^{(*)}$~\cite{Aaij:2020hsi} 
and $B_c \to J/\psi$~\cite{Aaij:2017tyk},
as well as in baryonic channels~\cite{Aaij:2015bfa,Detmold:2015aaa}, 
and increasingly precise $R$-ratio
determinations (see Fig.~\ref{fig:RX_projection_stat_syst}). 
Keeping pace with these advances is an important challenge
for the lattice community. Therefore now is a good time to take stock of
lattice efforts in these directions,
and this forms the main goal of the present article.

In the next section I briefly review the theory of semileptonic 
meson decays relevant for the direct determination of $|\Vcb|$.
The main component of this article is Sec.~\ref{LQCD-results} 
that attempts to summarise
the current status and works in progress on the lattice. Much of this material
is new/preliminary and was first presented at this conference.
Finally in Sec.~\ref{Conclusions} 
I will conclude with a short summary and some considerations for the future.

\section{Theory} \label{Theory}
The Standard Model parameter $|\Vcb|$ can be extracted precisely
using the semileptonic decay processes $B \to D^{(*)} l \nu_l$.
In these transitions the initial state $b$ quark is converted to
a $c$ quark by the weak interaction current, with an accompanying
factor of $\Vcb$ in the amplitude. In the Standard Model then
the differential partial widths for these decays are represented as follows:
\begin{align} \label{dGdw}
\frac{d \Gamma}{d w}(B \to D) &= (\text{known}) |\Vcb|^2 (w^2-1)^{3/2} 
|\mathcal{G}(w)|^2 \\
\frac{d \Gamma}{d w}(B \to D^*) &= (\text{known}) |\Vcb|^2 (w^2-1)^{1/2}
\chi(w) |\mathcal{F}(w)|^2
\end{align}
expressed here in terms of the kinematic variable $w$,
\begin{equation}
w = v_{B} \cdot v_{D^{(*)}} = 
\frac{M_B^2 + M_{D^{(*)}}^2 - q^2}{2 M_B M_{D^{(*)}}}
\end{equation}
Alternatively the kinematic variable $q^2$ is often used, where $q$
is the four-momentum transfer between initial and final state mesons.
In terms of these variables $q^2=0$ corresponds to maximum recoil
of the $D^{(*)}$ meson in the $B$ rest frame, while $w=1$ corresponds
to the $D^{(*)}$ at rest in the $B$ rest frame, or 
$q^2 = q^2_{\text{max}}= (M_B - M_D^{(*)})^2$.

In these expressions the non-perturbative QCD dynamics are contained
in the functions $\mathcal{F}(w)$ and $\mathcal{G}(w)$. In order to
determine $|\Vcb|$ from the experimental data involving these decays, these
functions need to be computed. The functions $\mathcal{F}(w)$ and
$\mathcal{G}(w)$ can in turn be expressed in terms of a number of form
factors, which are related to the following QCD matrix elements:

\begin{align}
\frac{\bra{D} V^\mu \ket{B}}{\sqrt{m_B \, m_D}} 
 &= (v_B + v_D)^\mu \green{h_+(w)} + (v_B -v_D)^\mu \green{h_-(w)} \\
\frac{\bra{D^*_\alpha} V^\mu \ket{B}}{\sqrt{m_B \, m_{D^*}}} 
&= \varepsilon^{\mu \nu \rho \sigma} 
v_B^\nu v_{D^*}^\rho \epsilon^{*\sigma}_\alpha \green{h_V(w)}\\
\frac{\bra{D^*_\alpha} A^\mu \ket{B}}{\sqrt{m_B \, m_{D^*}}} 
&= i \epsilon^{*\nu}_\alpha
\[ \green{h_{A_1}(w)} (1+w)g^{\mu \nu} - 
(\green{h_{A_2}(w)} v^\mu_B + \green{h_{A_3}(w)}v^\mu_{D^*}) v^\nu_B \]
\end{align}

These matrix elements can be computed from first principles using the
methods of lattice QCD, and from them the form factors determined.
In Sec.~\ref{LQCD-results} I will review the state-of-the-art in
these calculations, as well as for the related decays involving 
a $b \to c$ transition but with a strange or charm spectator chark. 
The formalism described above is analogous for these decays, 
but the form factors will differ. 

In the expressions~\eqref{dGdw}, there is a kinematic suppression factor
of $(w^2-1)$ raised to either the 3/2 or 1/2 power. This results in the 
experimental rates being damped near $w=1$, however, as will be discussed
in Sec.~\ref{LQCD-results}, most available lattice QCD results are limited
to the region $w \approx 1$. This is due to the fact that the lattice results
are more precise here, their signal decays at larger recoil. In addition,
the expressions for the rates~\eqref{dGdw} simplify at the zero-recoil point,
so that only a single form factor, $h_{A_1}(1)$ contributes.
Therefore the most precise determinations of $|V_{cb}|$ to date have focused
on precison lattice calculations of $h_{A_1}(1)$, 
combined with experimental data in $B \to D^*$, 
over a range of $w$ which is then extrapolated to $w=0$. 

In recent years some controversy has emerged regarding the precision 
with which one can reliably 
extract $|V_{cb}|$ using these extrapolations of the experimental data.
In this regard, different methods are now being utilised by both experimental
and lattice collaborations. While the
Caprini-Lellouch-Neubert (CLN)~\cite{Caprini:1997mu} uses an expansion
based on heavy quark effective theory valid to $\O(1/m_{b,c})$, the
Boyd-Grinstein-Lebed (BGL)~\cite{Boyd:1997kz}
is a model independent parameterisation based on analyticity
and unitarity.
The CLN approach has the
advantage of relying on few parameters, but this restrictiveness may
introduce model dependence particularly once a precision beyond
the level of approximation is reached~\cite{Bigi:2016mdz,Bordone:2019vic}.
The BGL approach is model independent and as a result relies on more parameters, 
and must be truncated at some order.

There have been several studies examining the model dependence
from different parameterisations in $B \to D$~\cite{Bigi:2016mdz} 
and $B \to D^*$~\cite{Bigi:2017njr,Grinstein:2017nlq,Dey:2019bgc} 
decays~\cite{Bernlochner:2017jka,Bernlochner:2017xyx}.
This progress was largely facilitated by experimental 
datasets with $q^2$ and angular distributions being made publicly available,  
including full error budgets and 
correlations~\cite{Abdesselam:2017kjf,Glattauer:2015teq}.
The situation was summarised recently by
FLAG~\cite{Aoki:2019cca}, reproduced in Fig.~\ref{fig:flag-1}, 
showing their best-fits for
$|\Vcb|$ utilising CLN and BGL parameterisations and compared with
the inclusive determination.

\begin{figure}[t]
\includegraphics[width=0.5\textwidth]{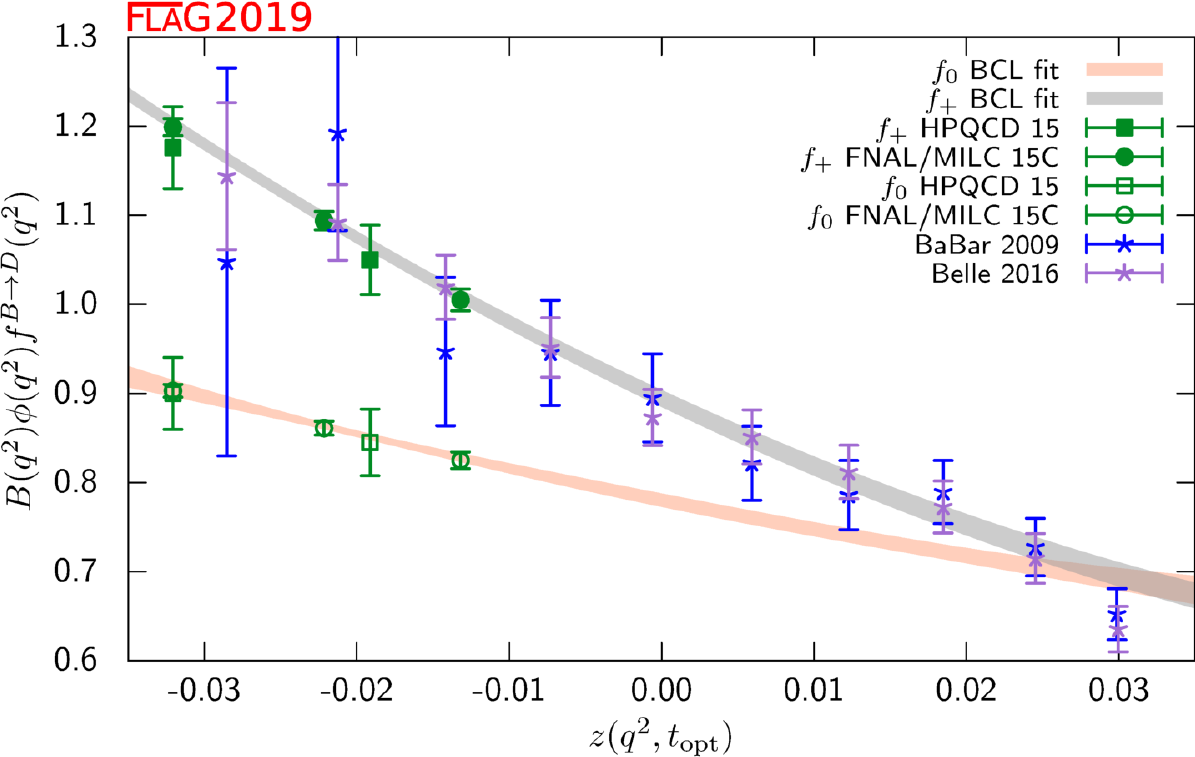} \hfill
\includegraphics[width=0.46\textwidth]{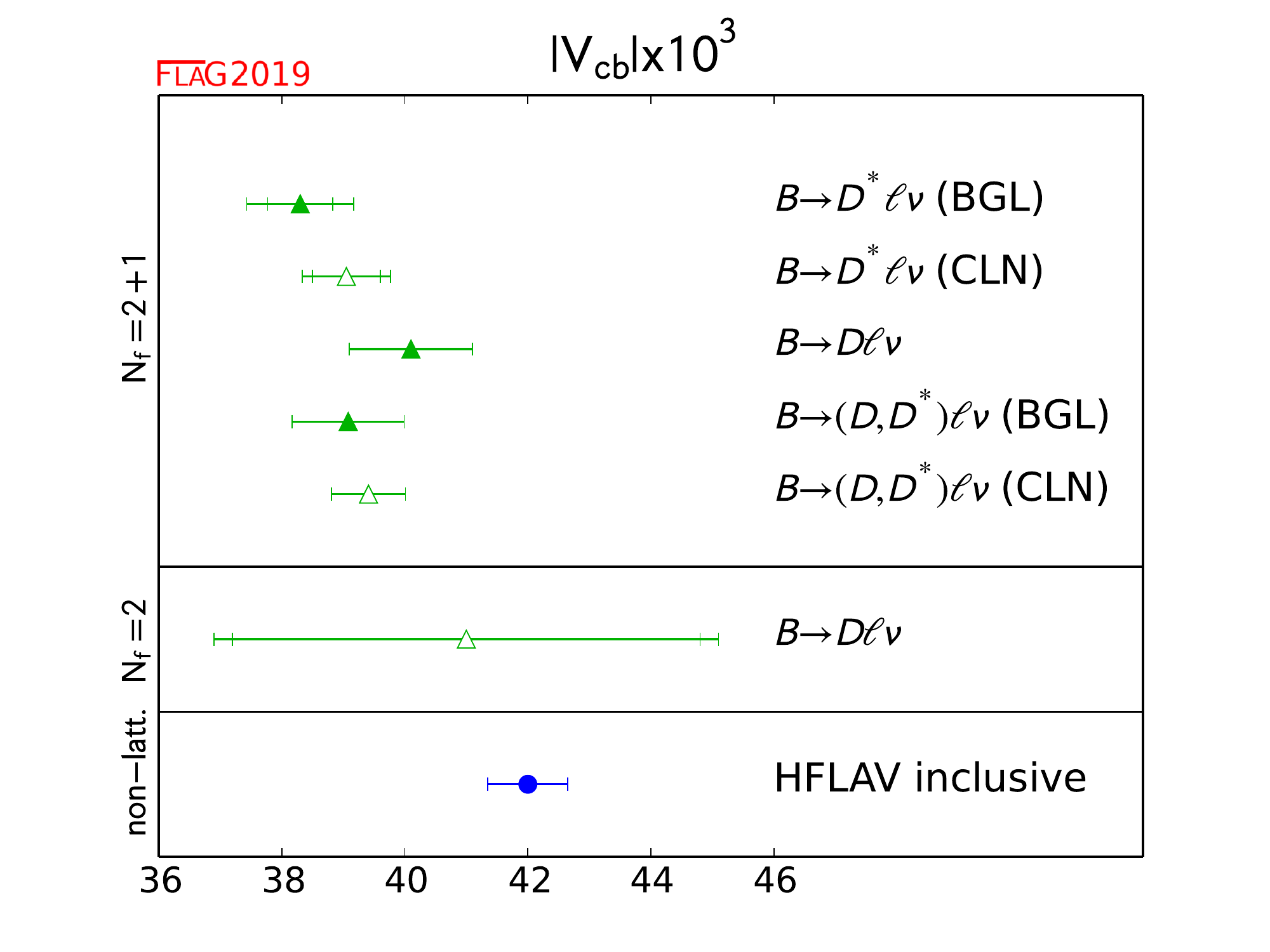}
\caption{(Left) Comparison of $B \to D$ semileptonic form factors as extracted
from experiment (blue, purple) and lattice QCD calculations (green),
as summarised in the 2019 FLAG review~\cite{Aoki:2019cca}. (Right)
Summary of different $|V_{cb}|$ extractions from FLAG varying the included decay
channels and parameterisations used.
\label{fig:flag-2}}
\end{figure}

In order to match experimental data with theory, it is also
extremely important for the 
lattice community to make predictions away from the
zero recoil point~\cite{Gambino:2019sif,Bernlochner:2017jka}. 
Interestingly, in the case of $B \to D$, the picture appears
somewhat more
congruent than for $B \to D^*$, and here form factors are available
over a large kinematic range both from 
experiment~\cite{Aubert:2009ac,Glattauer:2015teq} 
and lattice~\cite{Lattice:2015rga,Na:2015kha}. 
This is summarised in Fig.~\ref{fig:flag-2}. 
Although the final extraction of $|V_{cb}|^{\text{excl}}$
from this mode is less precise, it is also in reasonably good agreement
with the inclusive determination. With the expected improvements from
experiment and the lattice community, as more information becomes available, 
one imagines that the picture from $B \to D^*$ will become more clear.

It is also interesting to determine the related $B_s \to D_s^{(*)}$
form factors, which differ only in the substitution of the light spectator
quark for strange.
These decays were recently used by LHCb to measure $|\Vcb|$~\cite{Aaij:2020hsi}.
On the lattice, these calculations should be considerably less
computationally expensive due to the reduced cost of strange inversions
as compared to light, and also more statistically precise. Therefore
it provides both an interesting laboratory in which to test the effect of
different parameterizations, as well as make more precision checks between
competing lattice determinations. 
If there are systematic effects impacting a particular calculation of
$B \to D^*$, these should show up even more clearly in $B_s \to D^*_s$.
Therefore the channels $B_s \to D_s^{(*)}$ should be theoretically pursued. 

As will be discussed in Sec.~\ref{LQCD-results}, there
are now several efforts being undertaken by different groups
to extend these calculations beyond zero recoil, as well as 
explore other $b \to c$ decay modes so that the lattice can
be ready for the Belle II and new LHC eras. 

\section{Lattice QCD results} \label{LQCD-results}
Here I will briefly summarise the present status of lattice QCD calculations
for the group of semileptonic decays $B_{(s)} \to D_{(s)}^{(*)} \, l \nu$
and $B_c \to J/\psi \, l \nu$  , as well as planned efforts in these directions 
focusing on preliminary results presented in this conference.

One of the main features that distinguishes amongst 
these calculations is the choice
for the treatment of the $b$ quark in the simulation. Because $a m_b$ is 
not small for the lattice spacings used in many modern simulations, including
$b$ in the simulation on the same footing as the other quarks would lead
to uncontrollable lattice discretisation errors $\sim (am_b)^n$. Note
that the same considerations also hold for the charm quark, though it is
increasingly common to include it relativistically as lattice spacings 
decrease.
Therefore a strategy such as an effective theory framework
must be adopted to incorporate the $b$ quark.

Alternatively, at sufficiently small lattice spacings, it is possible
to work at masses $m_h$ approaching $m_b$ such that 
$a m_h \lesssim 1$. By working at several mass values approaching $m_b$,
the lattice data can be extrapolated to a physical prediction at $m_b$.
In what follows I will refer to this as a `relativistic-$b$' approach
to distinguish it from the effective treatments of $b$. 
This method
benefits from the use of improved actions which formally improve
the form of leading heavy mass discretisation effects, 
as used in the `heavy-HISQ'~\cite{McNeile:2012qf} results shown later.
A feature of the relativistic-$b$ approach is that
it gives not only predictions for the physical decay of a
$B$ meson, but also results for masses of the heavy quark
between $m_c$ and $m_b$. 

\subsection{$B \to D^*$}

\begin{figure}[t]
\includegraphics[width=0.44\textwidth]{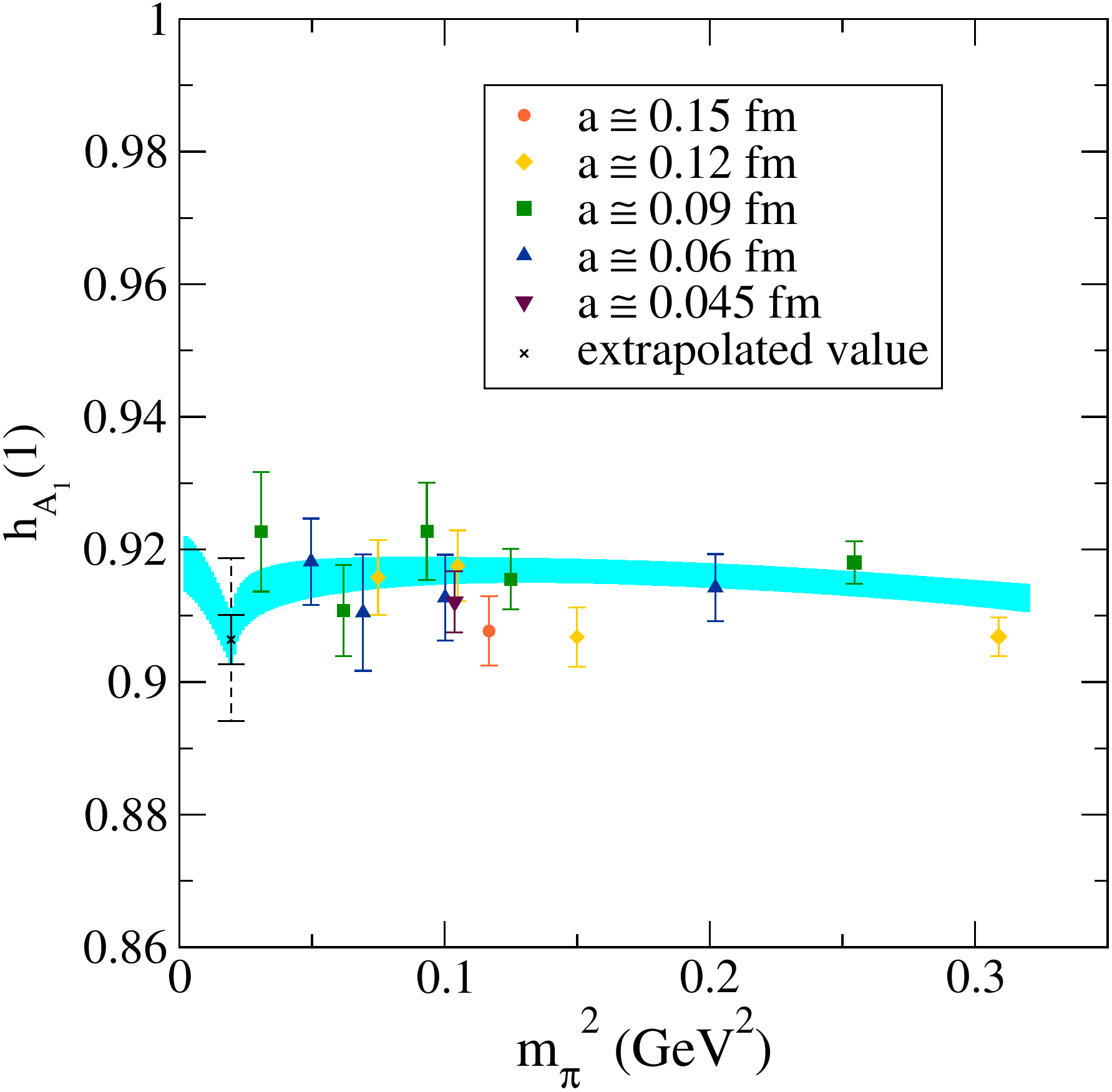} \hfill
\includegraphics[width=0.5\textwidth]{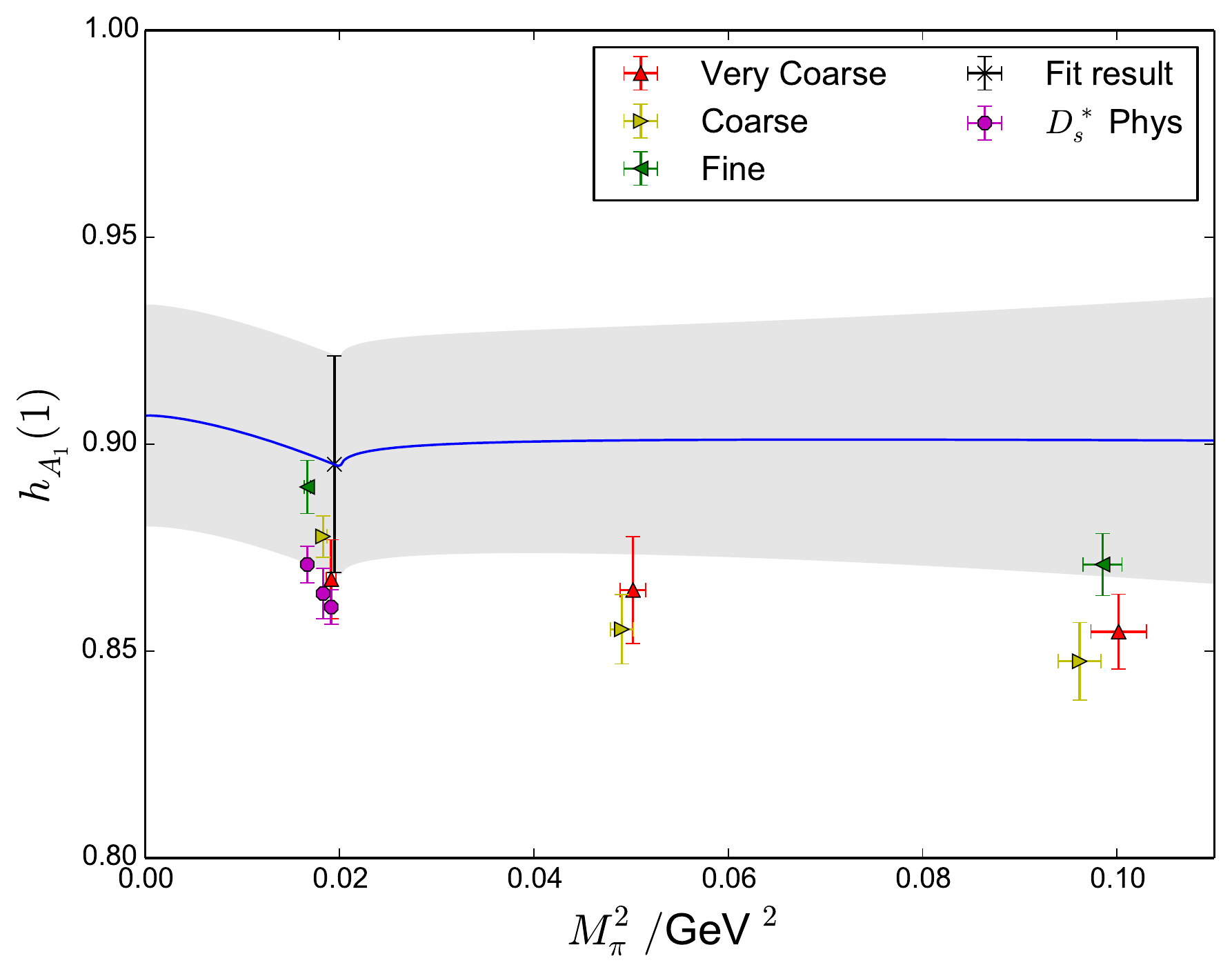}
\caption{
Comparison of chiral-continuum extrapolations for the 
$B \to D^*$ form factor $h_{A_1}$
at the zero-recoil point, computed by the 
FNAL-MILC~\cite{Bailey:2014tva} (left) and HPQCD~\cite{Harrison:2017fmw} 
(right) collaborations. The cusp near the physical
$m_\pi^2$ comes from expectations of chiral perturbation theory.
\label{fig:B2Dstar_q2max}}
\end{figure}

There are two recent published lattice calculations 
of the $B \to D^* \, l \nu$ decay, both at the zero-recoil point, 
where only the single form factor $h_{A_1}$ 
contributes. One is from the Fermilab/MILC collaboration, calculated
on $n_f = 2+1$ MILC asqtad ensembles, using clover heavy quarks with the
Fermilab interpretation~\cite{Bailey:2014tva}.
This calculation makes use 
of the `ratio-method'~\cite{Bernard:2008dn,Hashimoto:2001nb,Hashimoto:1999yp},
wherein they calculate a specific double ratio
\begin{equation}
\frac{
\bra{D^*} \bar{c} \gamma_j \gamma_5 b \ket{\overline{B}} 
\bra{\overline{B}} \bar{b} \gamma_j \gamma_5 c\ket{D^*}}
{\bra{D^*}\bar{c} \gamma_4 c \ket{D^*} 
\bra{\overline{B}}\bar{b} \gamma_4 b \ket{\overline{B}}} = 
|h_{A_1}(1)|^2 \,,
\end{equation}
to cancel systematic and statistical errors. 

The other is from HPQCD collaboration on $n_f=2+1+1$ MILC HISQ ensembles,
using non-relativistic QCD for the $b$-quark~\cite{Harrison:2017fmw}. 
The chiral extrapolations for the quantity $h_{A_1}(1)$
are compared in Fig.~\ref{fig:B2Dstar_q2max}. 
The two groups find compatible results,
quoting $h_{A_1}(1) = 0.906(4)(12)$ and 0.895(10)(24) for FNAL/MILC and
HPQCD respectively. The dominant error in~\cite{Harrison:2017fmw} arises
from missing $\O(\alpha_s^2)$ matching of NRQCD currents to QCD.

\begin{figure}[t]
\begin{subfigure}{0.5\textwidth}
\includegraphics[width=0.45\textwidth]{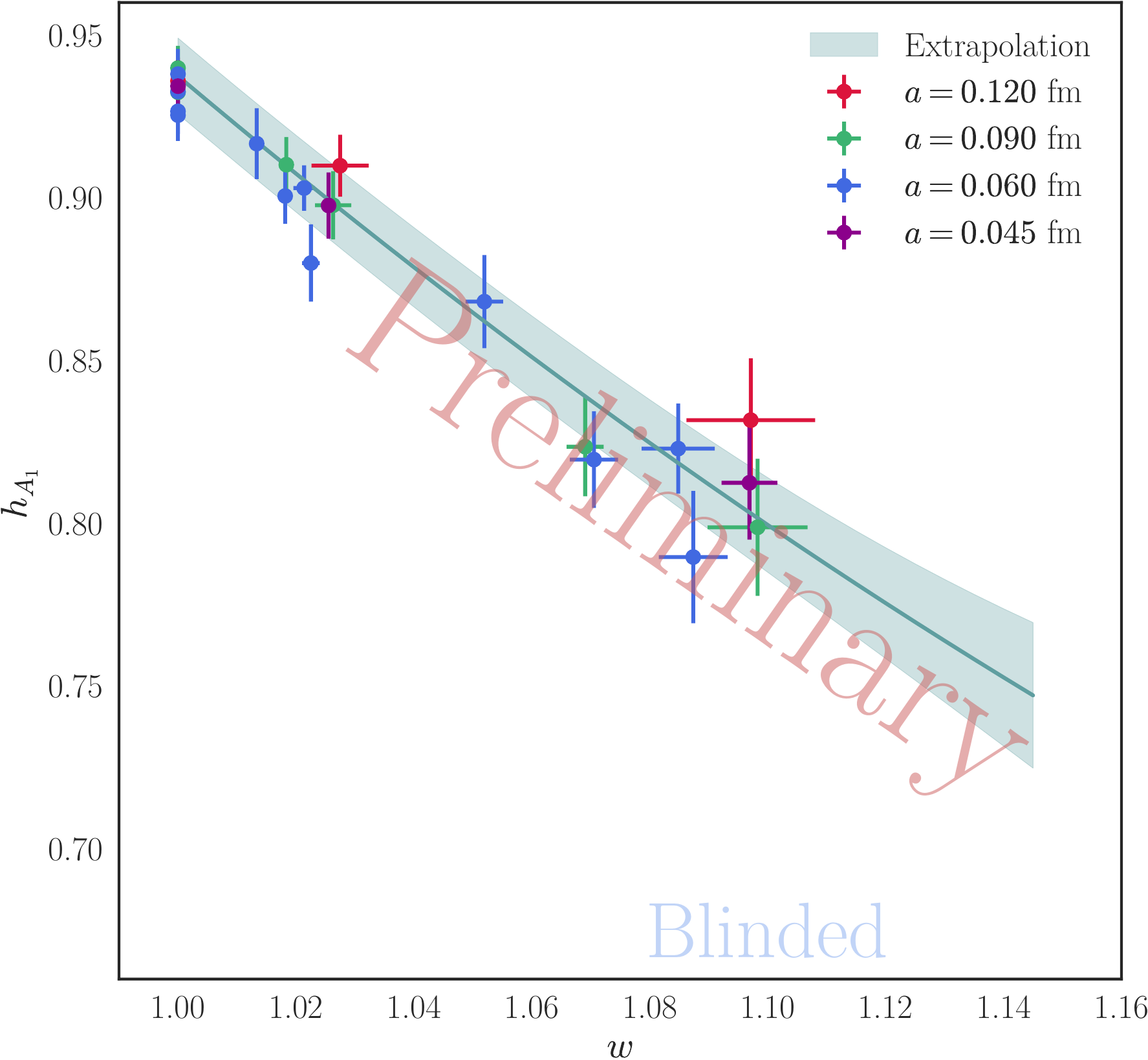} \hskip 0.05cm
\includegraphics[width=0.45\textwidth]{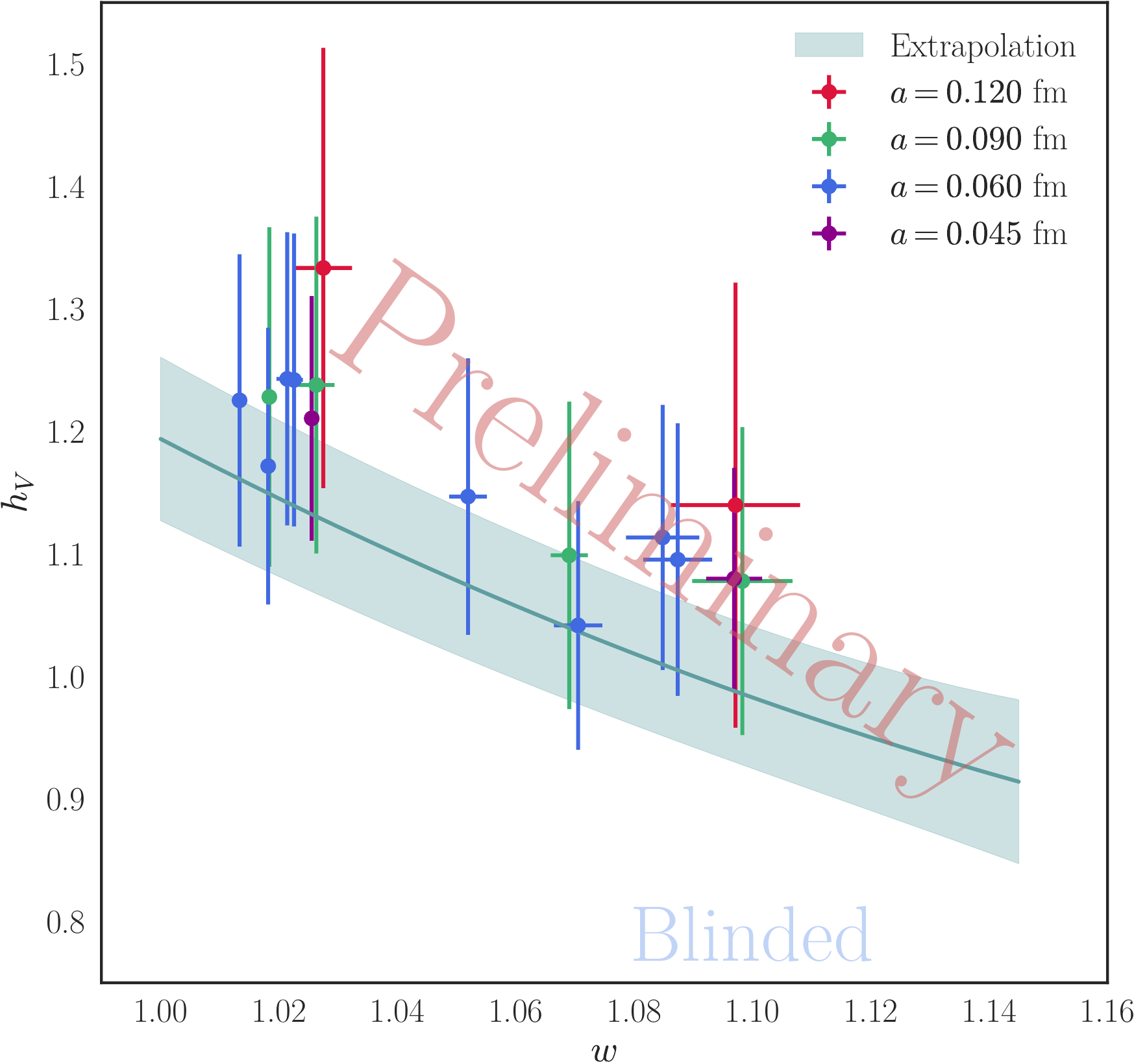} \\ 
\includegraphics[width=0.45\textwidth]{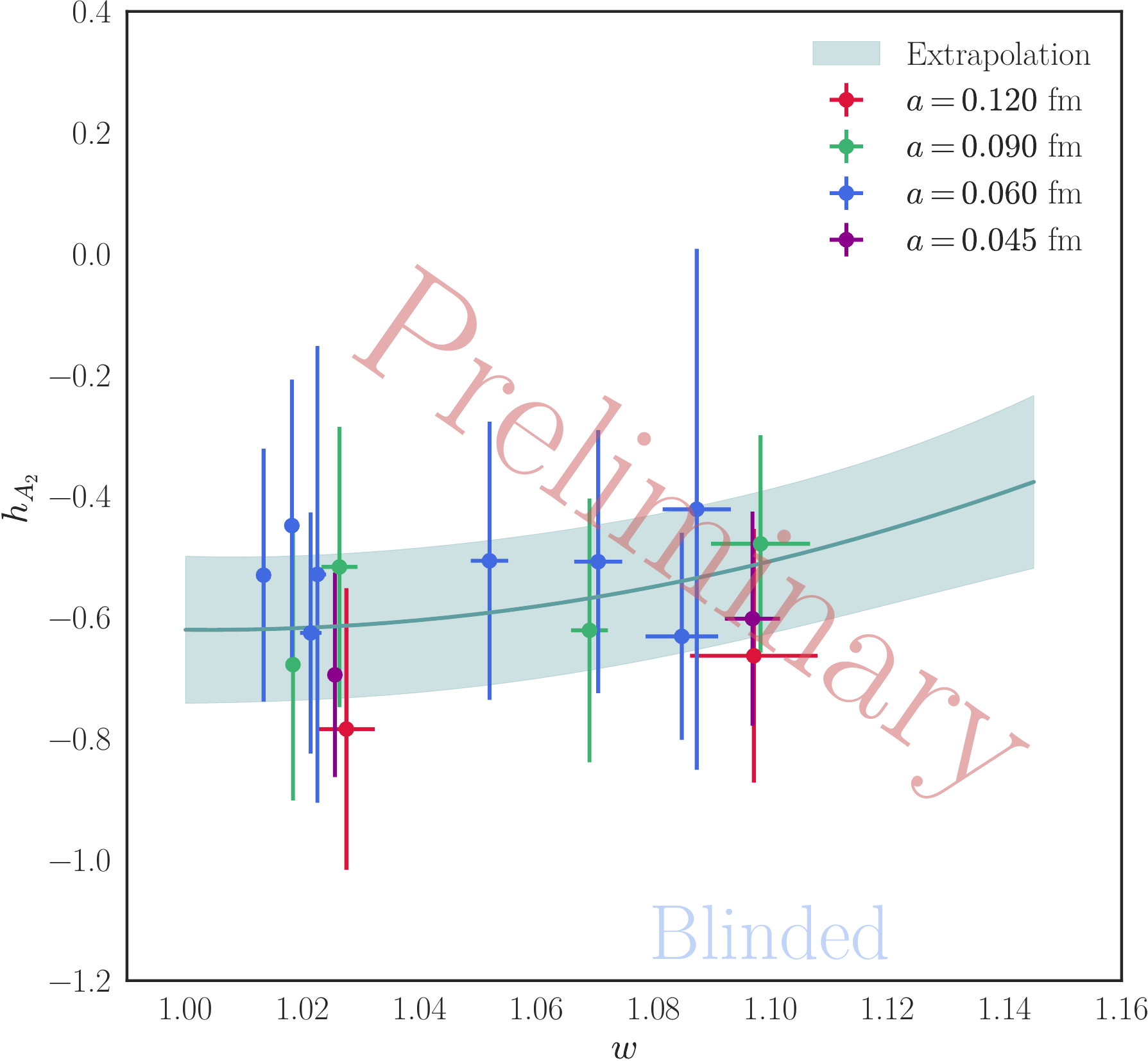} \hskip 0.05cm
\includegraphics[width=0.45\textwidth]{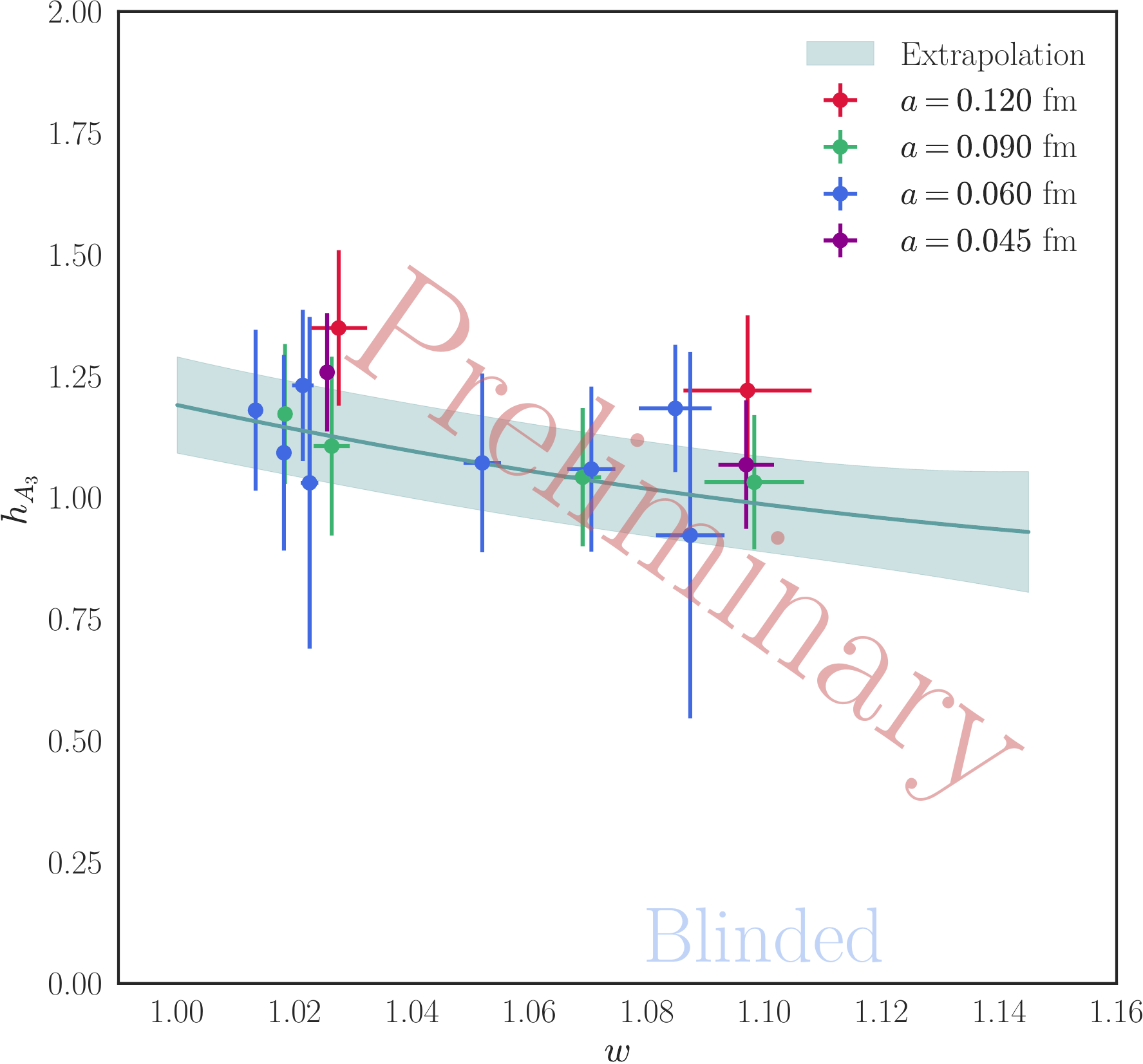}\\
\end{subfigure}
\begin{subfigure}{0.5\textwidth}
\includegraphics[width=0.9\textwidth]{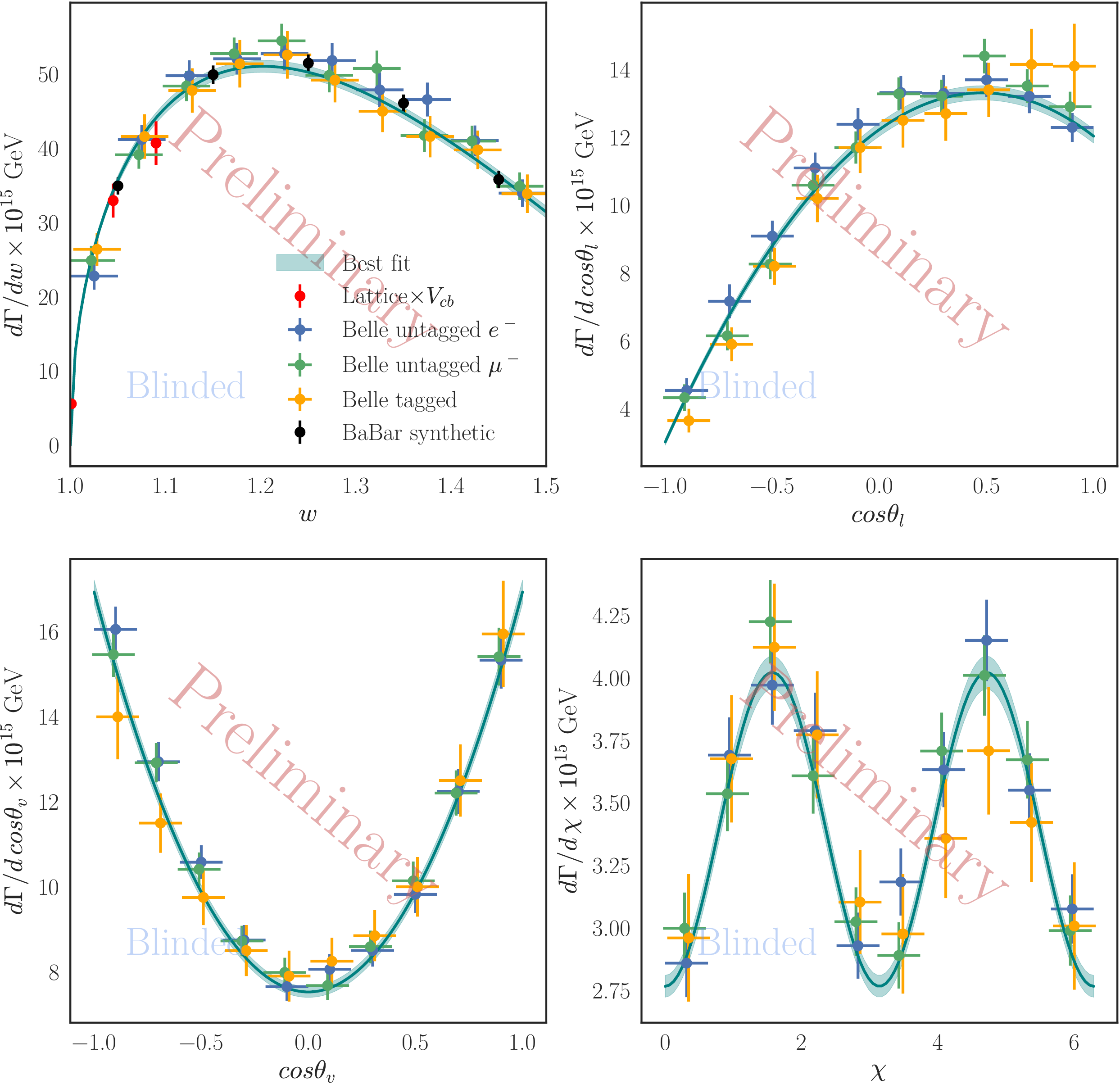} \hfill
\end{subfigure}
\caption{$B \to D^*$ form factors from the MILC collaboration.
The figures on the left show the lattice form factors at different
values of the lattice spacing and the extrapolated results, the figures
on the right show the kinematic distributions
derived from a combination fit of lattice and experimental data
compared with lattice data and data from Belle and Babar.
See also results in~\cite{Aviles-Casco:2017nge,Vaquero:2019ary,Vaquero-latt19}
Figs.\ courtesy A.\ Vaquero.
\label{fig:milc_update}}
\end{figure}

The MILC collaboration is extending their calculation to the the full set of
form factors, away from zero recoil.
Preliminary results in the range $w \in [1, 1.1]$ were presented
in~\cite{Aviles-Casco:2017nge}, and an update was presented at this 
conference~\cite{Aviles-Casco:2019zop} showing global fits and comparison 
with available experimental data, 
as shown in Fig.~\ref{fig:milc_update}.

\begin{figure}[t]
\includegraphics[width=0.44\textwidth]{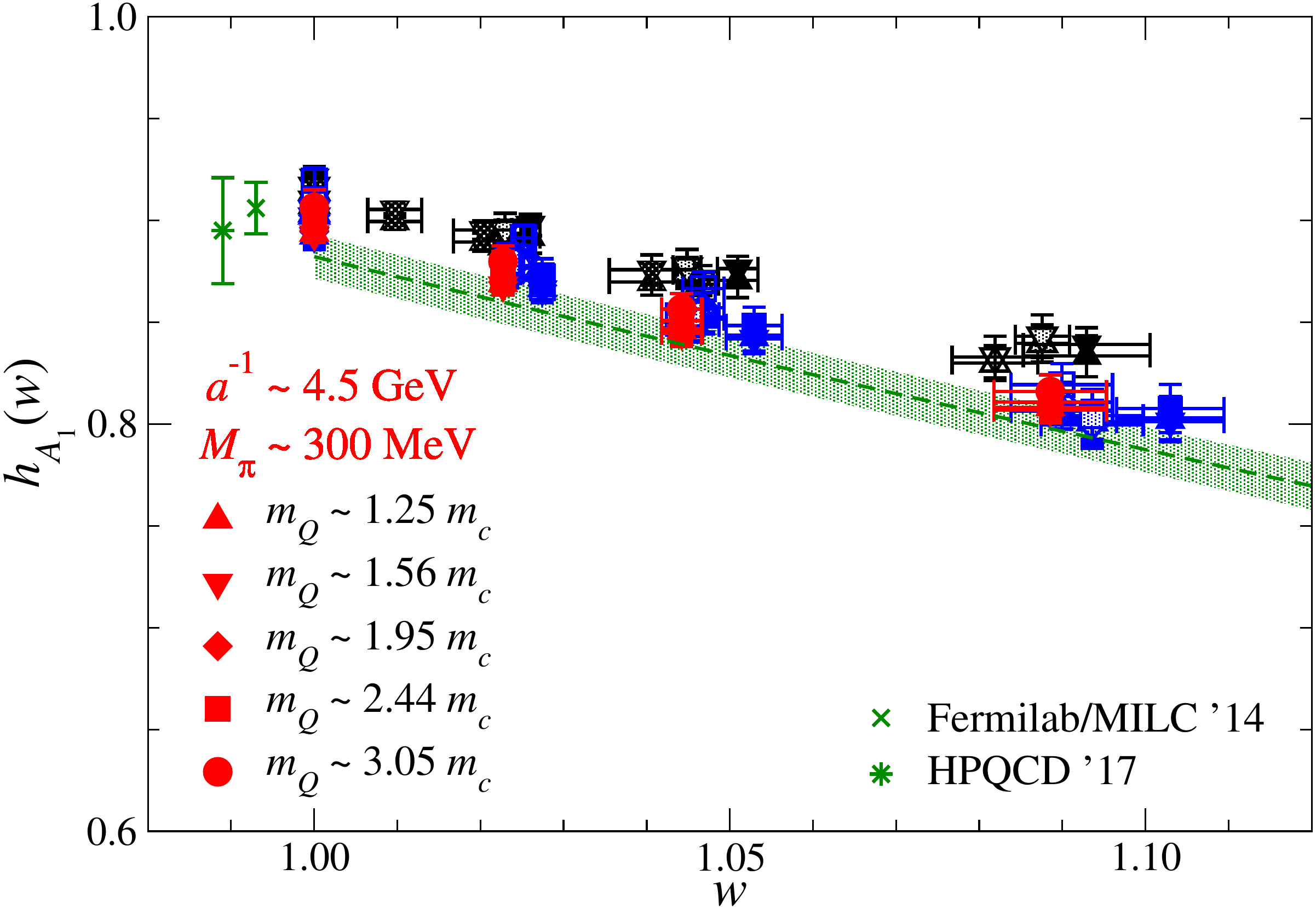} \hfill
\includegraphics[width=0.44\textwidth]{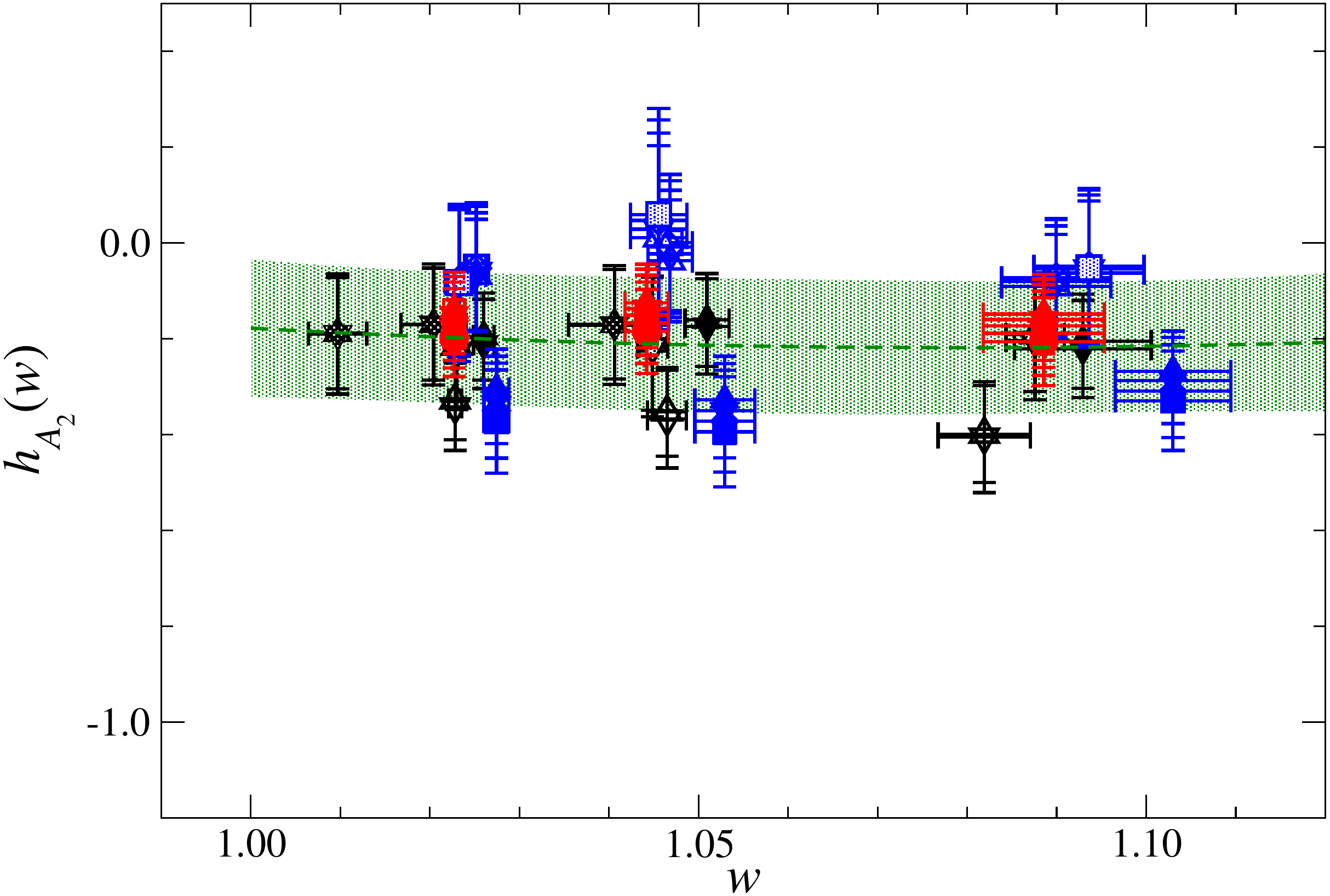}
\vskip0.4cm
\includegraphics[width=0.44\textwidth]{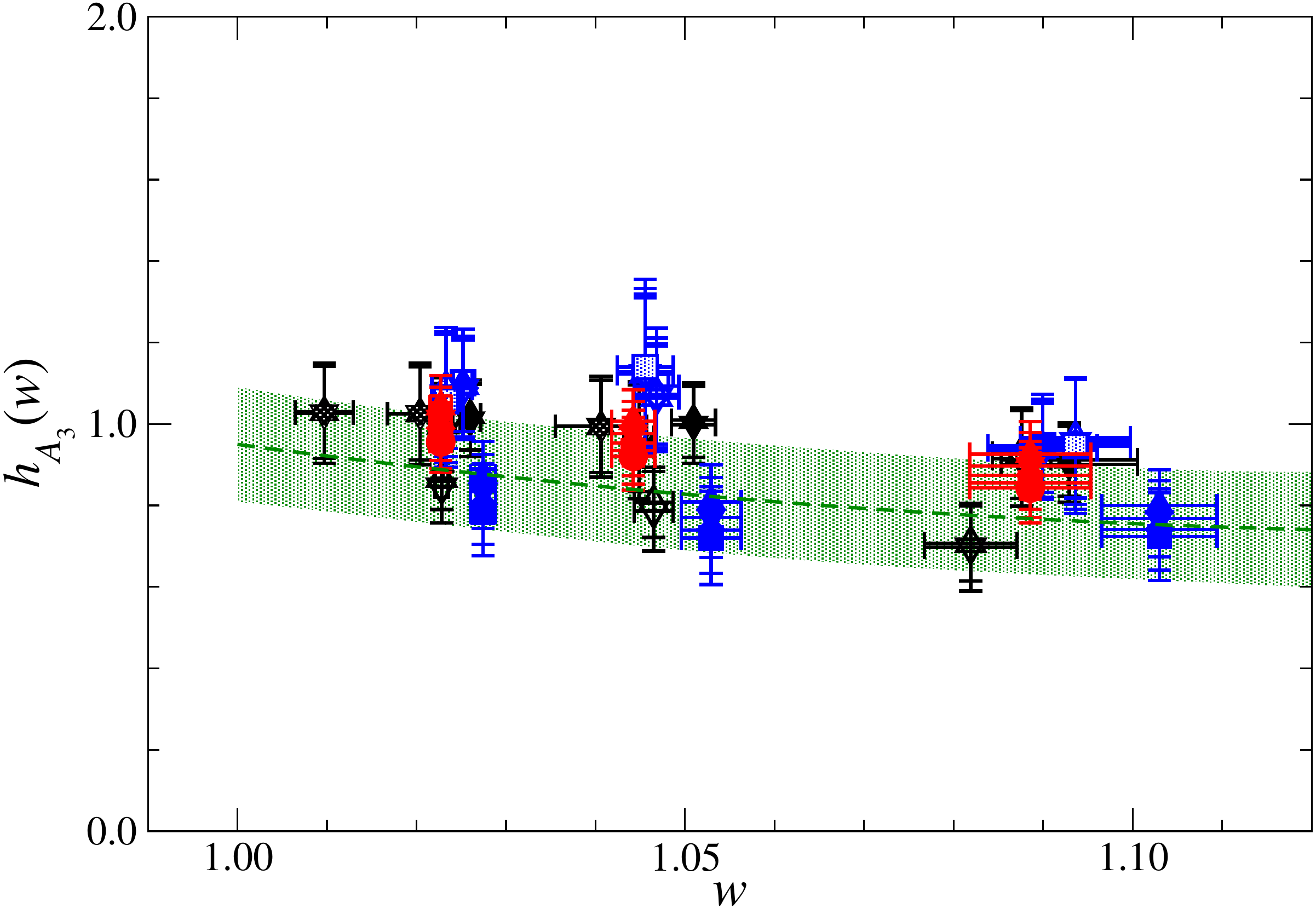} \hfill
\includegraphics[width=0.44\textwidth]{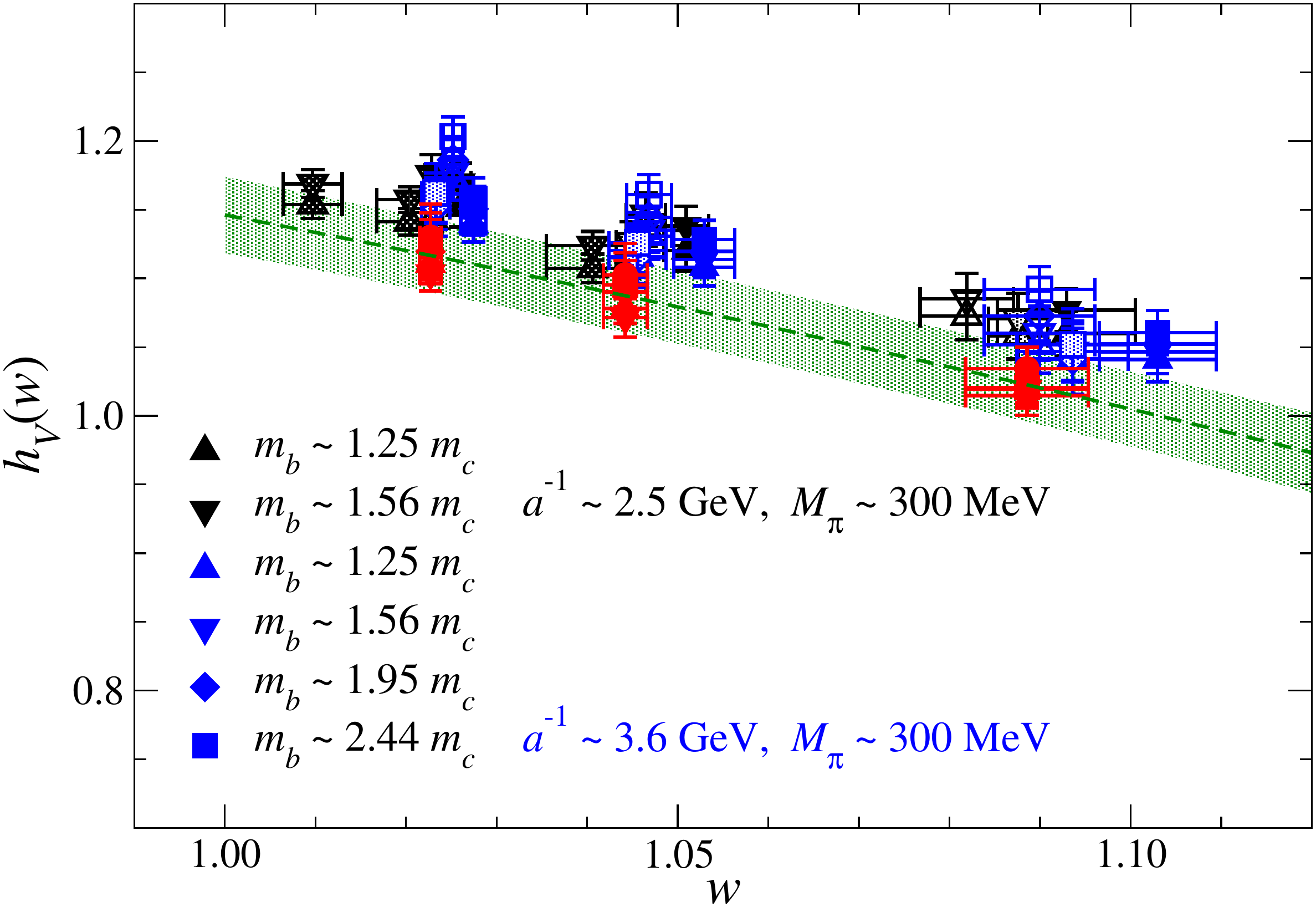} \\
\caption{$B \to D^*$ form factors from the JLQCD collaboration,
using a `relativistic-$b$' approach. New results on a finer 
$a^{-1} \approx$ 4.5 GeV lattice spacing are shown in red, and
the result of chiral/continuum extrapolations are shown as green
bands. For $h_{A_1}(1)$ the comparison with previous 
results from~\cite{Bailey:2014tva,Harrison:2017fmw} are also shown.
Figs.~courtesy Takashi Kaneko. 
\label{fig:B2Dstar_jlqcd}}
\end{figure}

The JLQCD collaboration have presented their preliminary results
for $B \to D^{(*)}$ form factors in~\cite{Kaneko:2018mcr} for a range
$w \in [1, 1.06]$ and at two lattice spacings, 
and an update of these results were presented at
this conference by Kaneko~\cite{Kaneko:2019vkx}, with extended range
in $w \in [1, 1.1]$ and including results at a finer lattice spacing
of $a^{-1} = 4.5$ GeV. These calculations use a `relativistic-$b$'
approach on M\"obius domain wall fermions, for which
observables are calculated over a range of heavy quark
masses keeping $am_h < 0.8$, ($m_h$ up to 3.05 $m_c$), with an extrapolation
required to the physical $m_b$. Their results are shown in 
Fig.~\ref{fig:B2Dstar_jlqcd} for $h_{A_{1,2,3}}$ and $h_{V}$. 
The extrapolated results for $h_{A_1}(1)$ agree well 
with FNAL/MILC and HPQCD~\cite{Bailey:2014tva,Harrison:2017fmw}.

\begin{figure}[t]
\includegraphics[width=0.45\textwidth]{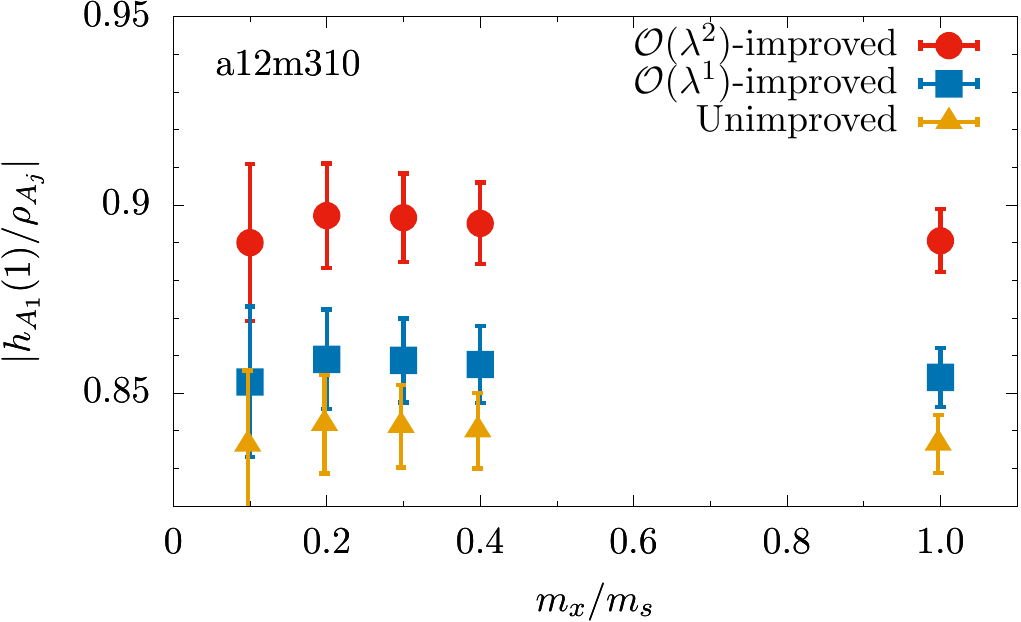} \hfill
\includegraphics[width=0.45\textwidth]{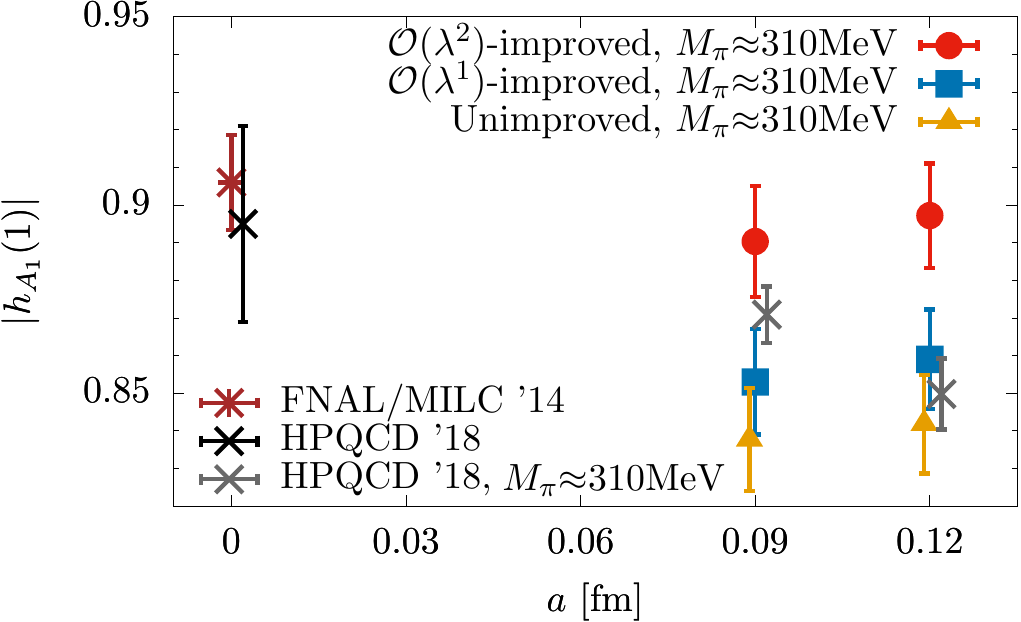}
\caption{Results for $B \to D^*$ at zero recoil 
from LANL/SWME~\cite{Bhattacharya:2018ibo} showing the effect of
higher orders of current improvement (left) 
and compared with prior results in the literature (right) --
Note that $\rho_{A_1}$ is here set to 1 so the comparison is only
indicative.
\label{fig:BtoDstar_lanl}}
\end{figure}

The LANL/SWME collaboration have also released preliminary results for the
$h_{A_1}$ form factor at zero 
recoil~\cite{Bailey:2017xjk,Bhattacharya:2018ibo}, 
and at this conference~\cite{Bhattacharya:2020xyb}.
Their calculation is carried out on the $n_f=2+1+1$ MILC HISQ
ensembles, using the 
Oktay-Kronfeld (OK) action~\cite{Oktay:2008ex,Bailey:2017nzm} 
for valence charm and bottom, at two lattice spacings 
$a \approx 0.12, 0.09$ fm and a single pion mass $m_\pi \approx 310$ MeV.
The OK action is improved at a higher order in 
$\lambda_{c,b} \sim \frac{\Lambda_{\text{QCD}}}{2 m_{c,b}}$ 
than the Fermilab action being used by the MILC collaboration --
This is important to reduce the charm
quark discretisation error, the dominant (1\%) error
in~\cite{Bailey:2014tva}, to below the percent level~\cite{Bailey:2020uon}.
Preliminary results for $h_{A_1}(1)$ are shown in Fig.~\ref{fig:BtoDstar_lanl}. 

\subsection{$B \to D$}
The FNAL/MILC and HPQCD collaborations have both computed $B \to D$
form factors at zero and non-zero recoil on $n_f = 2 + 1$ MILC asqtad
lattices, the former using heavy quarks in the Fermilab
approach~\cite{Lattice:2015rga} and with lattice spacing down to
$a \sim 0.045$ fm,
the latter using HISQ (relativistic) $c$ and NRQCD $b$~\cite{Na:2015kha}
at two lattice spacings of $a \sim 0.09$, 0.12 fm. A comparison
of these results reproduced from~\cite{Aoki:2019cca} 
is shown in Fig.~\ref{fig:flag-2} along with experimental
data~\cite{Aubert:2009ac, Glattauer:2015teq}.
Their results
are in good agreement, although HPQCD has larger errors coming mainly from
discretization effects and
NRQCD matching uncertainties, 
similar to the situation for $B \to D^*$. 

In contrast to the present situation with $B \to D^*$, here the form
factors from lattice are available over an extended range in $q^2$.
After the new lattice data beyond zero-recoil became available as well as
new experimental data from Belle, a careful analysis~\cite{Bigi:2016mdz} 
of the available data and 
different form factor parameterisations found a value
for $|V_{cb}| = 40.49(97)10^{-3}$, this value being between, and 
compatible with, both the inclusive determination
and the exclusive value from $B \to D^*$. 
It is clear from this study the
importance of having lattice data away from zero recoil, as well
as carefully assessing parameterisation dependence.
The lattice and experimental data for the form factors are shown 
together in Fig.~\ref{fig:flag-2}.

\begin{figure}[t]
\centering
\includegraphics[width=0.5\textwidth]{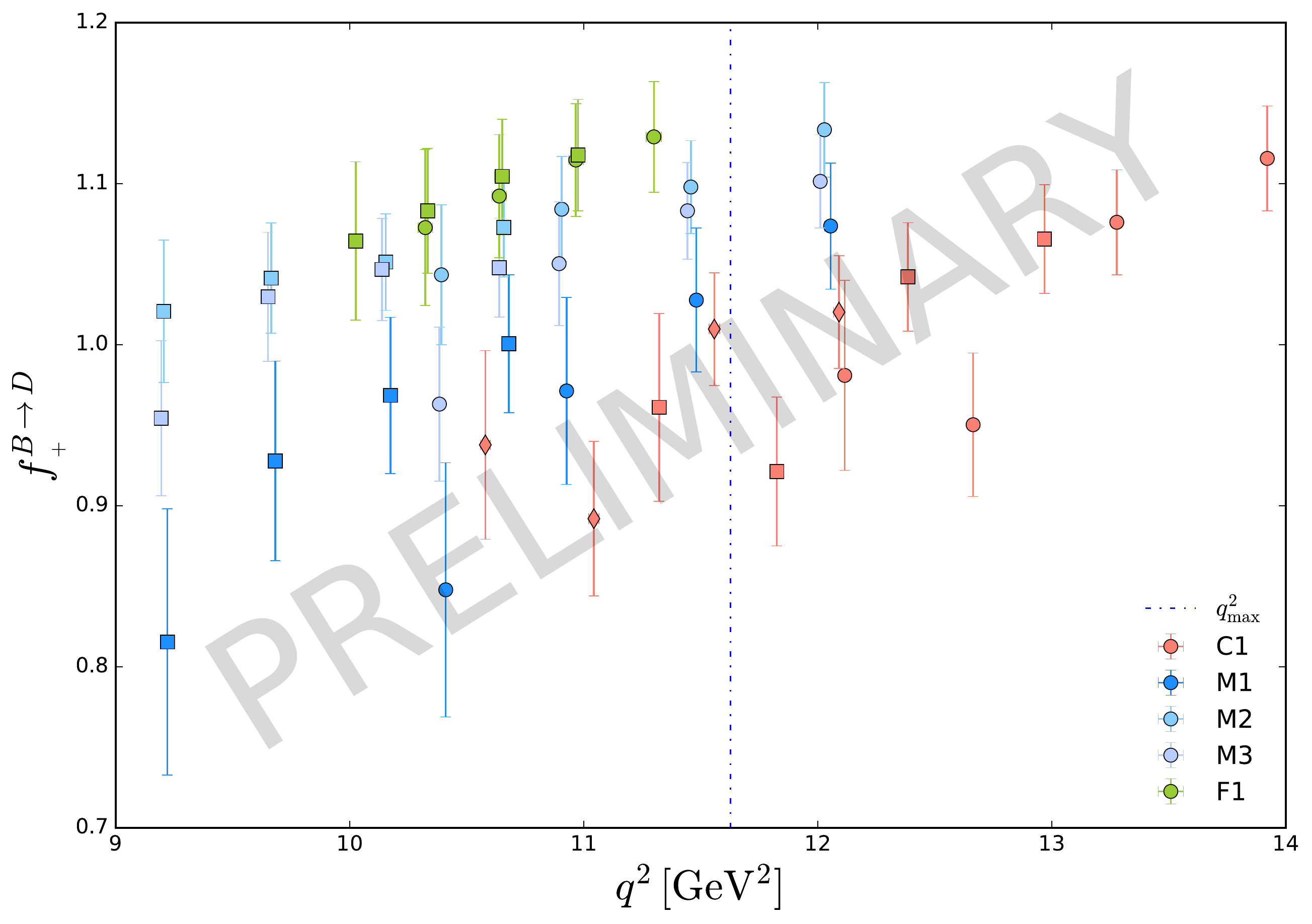} \hfill
\includegraphics[width=0.45\textwidth]{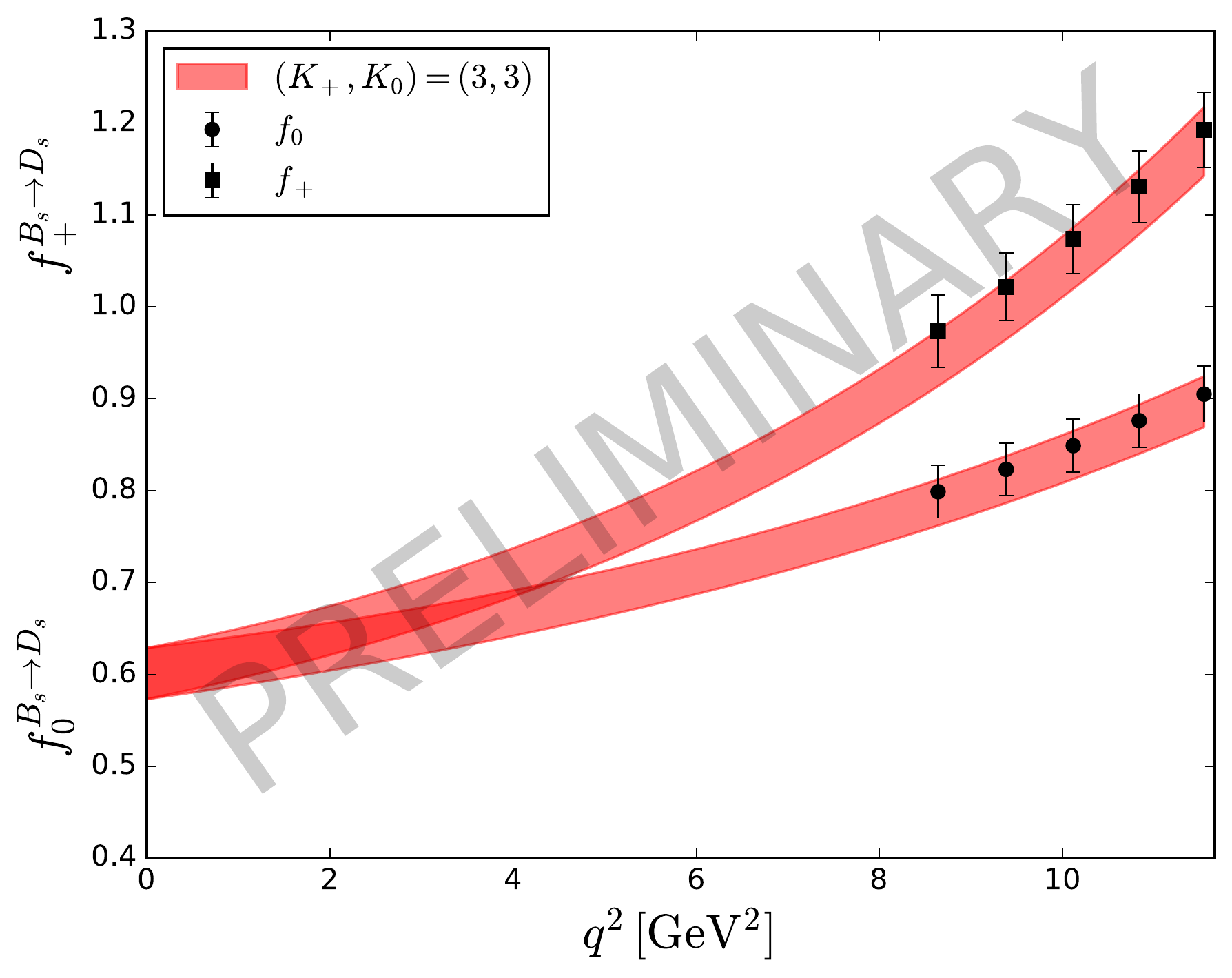}
\caption{
Preliminary results from the RBC/UKQCD collaboration~\cite{Flynn:2019jbg}.
(Left) Data for $f_+(q^2)$ for $B \to D$ decay. The colors correspond to 
the different gauge ensembles ($a$/$L$/$m_\pi$), 
and different shapes correspond with
different input "charm" masses used to interpolate/extrapolate to 
physical charm.
(Right) The $f_{+/0}(q^2)$ form factors for $B_s \to D_s$
with fits to a CLN parameterization.
\label{fig:BstoDs_rbc}}
\end{figure}

\begin{figure}[t]
\includegraphics[width=0.49\textwidth]{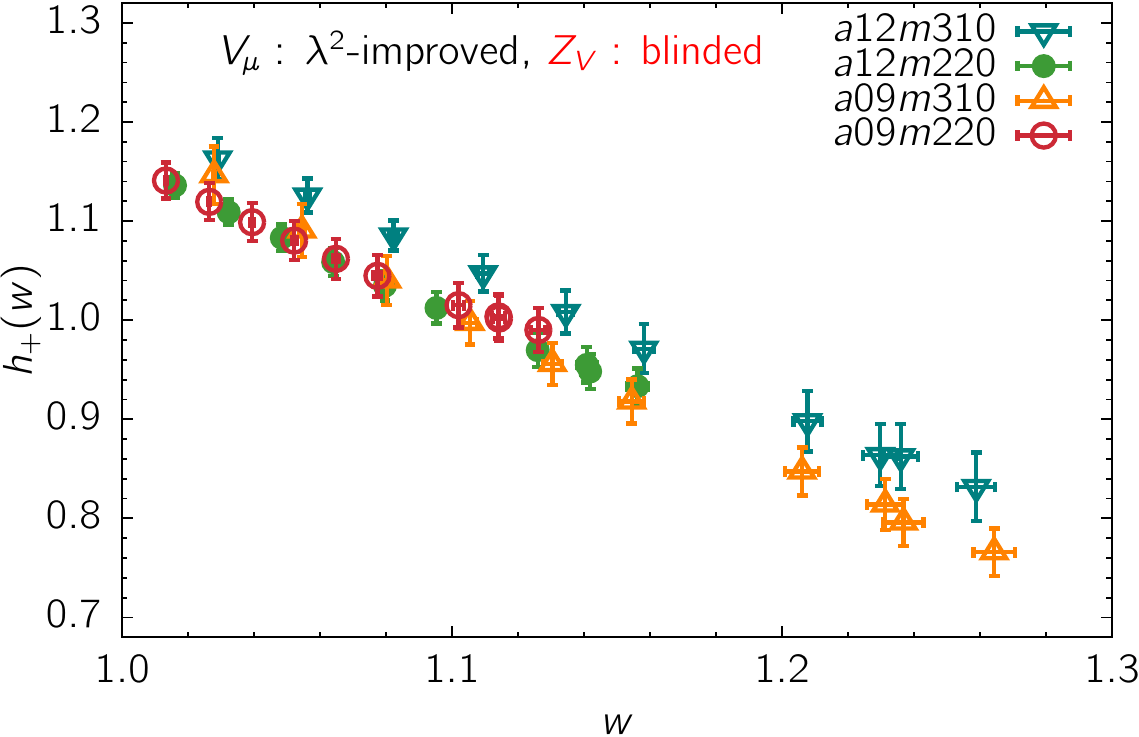}
\includegraphics[width=0.49\textwidth]{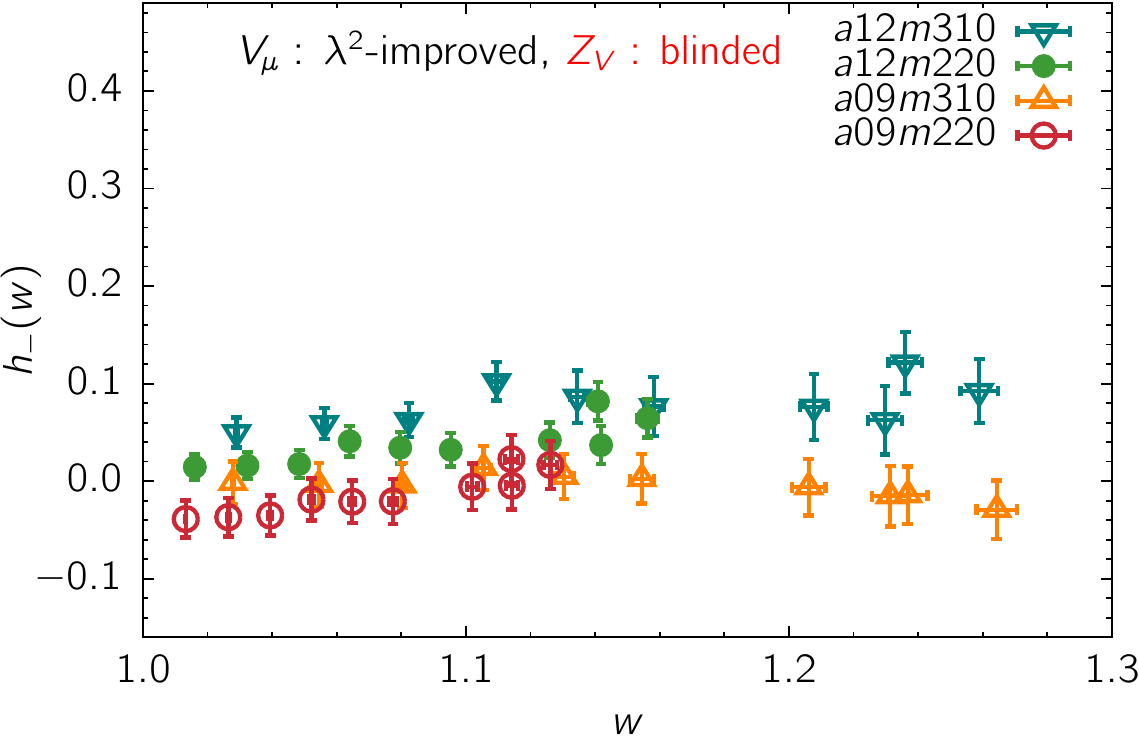}
\caption{Preliminary results for $B \to D$ $h_{+}(w)$ (left) 
and $h_{-}(w)$ (right) form factors
from LANL/SWME~\cite{Bhattacharya:2020xyb}, at two lattice spacings
($a \approx 0.12, 0.09$ fm) and two pion masses ($m_\pi \approx$ 310, 220 MeV). 
The overall normalisation of the results is blinded.
\label{fig:BtoD_lanl}}
\end{figure}

Preliminary results for $B \to D$ form factors from JLQCD collaboration
were presented in~\cite{Kaneko:2018mcr}, with an update presented
at this conference~\cite{Kaneko:2019vkx} including lighter
pion masses, and a third lattice
spacing with $a^{-1} \sim 4.5$ GeV, which allows to extend
the heavy quark mass in the simulation to $m_h \sim 3.05 m_c$.
RBC/UKQCD also presented~\cite{Flynn:2019jbg} preliminary results on 
$n_f=2+1$ domain wall ensembles,
treating light, strange, and charm quarks with the domain wall action,
and the bottom quark with a 
relativistic heavy quark action as shown
in Fig.~\ref{fig:BstoDs_rbc}. Preliminary results
at two lattice spacings
($a \approx 0.12, 0.09$ fm) and two pion masses ($m_\pi \approx$ 310, 220 MeV)
were presented by LANL/SWME~\cite{Bhattacharya:2020xyb},
their results for the $h_{+/-}(w)$ form factors are shown in
Fig.~\ref{fig:BtoD_lanl}.

\begin{figure}[t]
\includegraphics[width=0.46\textwidth]{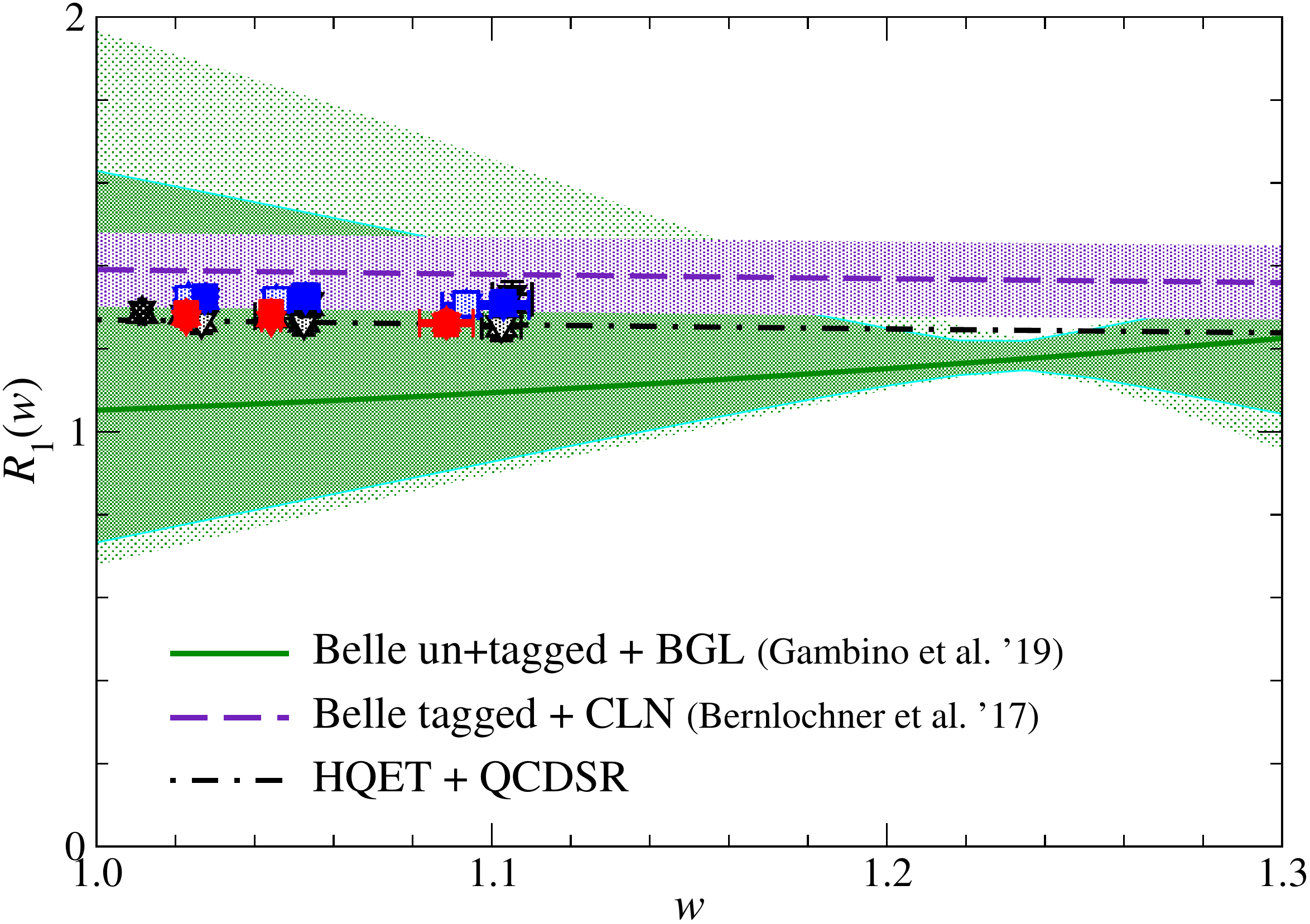} \hfill
\includegraphics[width=0.46\textwidth]{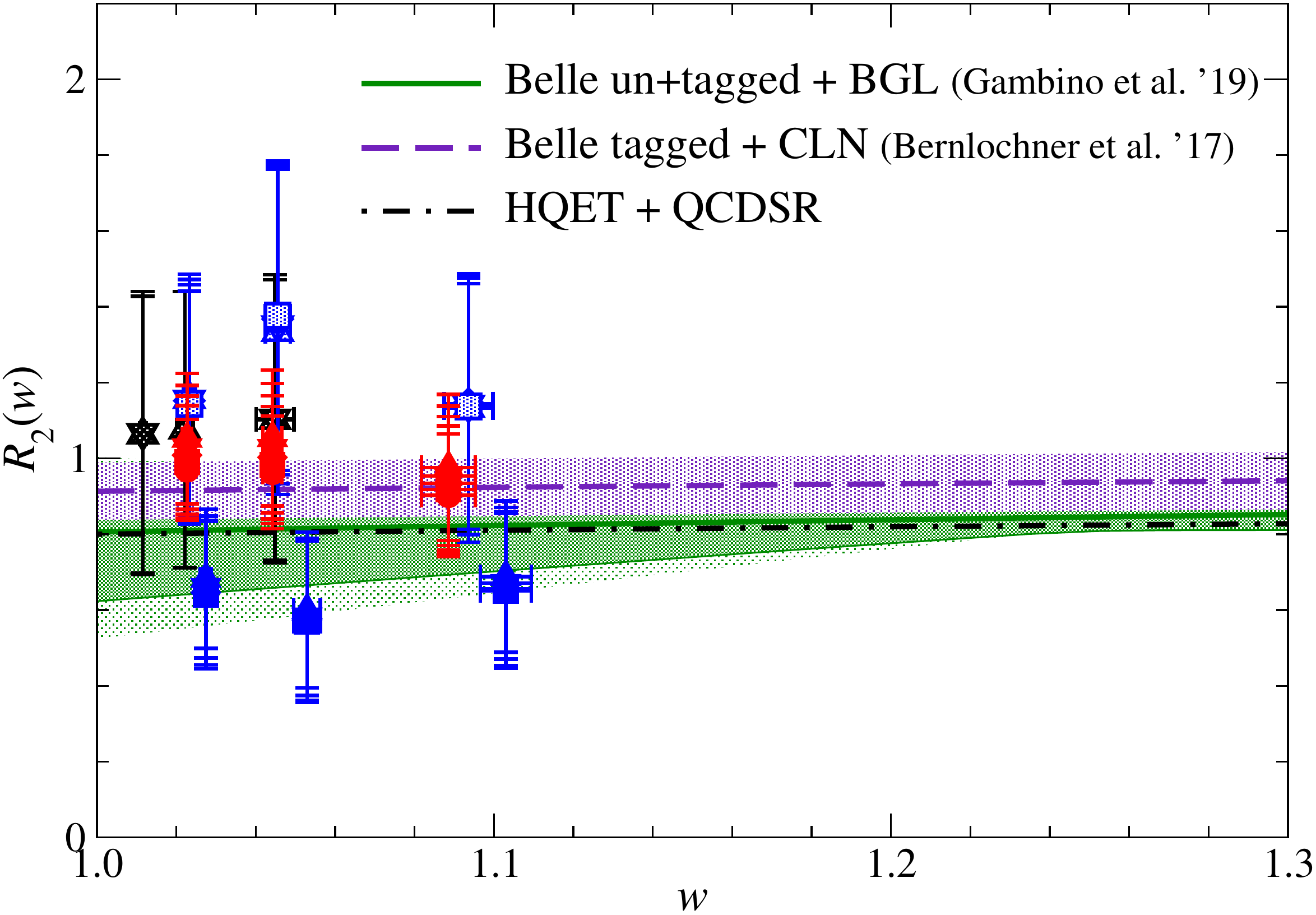}
\vskip 0.75cm
\caption{Comparison of lattice form factor data from JLQCD (symbols)
with fits to experimental Belle
data~\cite{Abdesselam:2017kjf,Abdesselam:2018nnh} 
using different parameterisations 
(colored bands)~\cite{Bernlochner:2017xyx,Gambino:2019sif}
and predictions of HQET. 
For more discussion see~\cite{Kaneko:2019vkx}.
Figs.\ courtesy Takashi Kaneko. 
\label{fig:R1R2_jlqcd}}
\end{figure}

\subsection{$B_s \to D^*_s$}
There are two determinations of the $B_s \to D_s^*$ zero-recoil
form factor $h_{A_1}^s(1)$, both from the HPQCD collaboration using
$n_f=2+1+1$ MILC HISQ ensembles, 
but differing in the treatment of the $b$-quark.
The calculation of~\cite{Harrison:2017fmw} uses an NRQCD $b$-quark 
on relatively coarser
ensembles, while~\cite{McLean:2019sds} uses the relativistic 
`heavy-HISQ' approach on
fine ensembles down to $a \sim 0.45$ fm. The main systematic uncertainty
in the NRQCD calculation comes from the perturbative current matching
known to $\O(\alpha_s)$,  this error is absent 
from the heavy-HISQ calculation where the current
is normalised non-perturbatively using the PCAC relation.
The two calculations are in agreement
\begin{align}
h^s_{A_1}(1) &= 0.883(12)_{\text{stat}} (28)_{\text{sys}} \\
h^s_{A_1}(1) &= 0.9020(96)_{\text{stat}} (90)_{\text{sys}}
\end{align}

It is also interesting to note that in~\cite{Harrison:2017fmw} 
the ratio of zero-recoil form factor with light/strange spectator 
was calculated to be
$h_{A_1}(1)/h^s_{A_1}(1) = 1.013(14)_\text{stat}(17)_\text{sys}$.
In this ratio the main systematic from the current matching largely cancels.

\subsection{$B_s \to D_s$}
There have been a few calculations of the $B_s \to D_s$ form factors,
using different methodologies.
The MILC collaboration determined $f_0(q^2)$ and $f_+(q^2)$
using $n_f=2+1$ asqtad ensembles, with charm and bottom valence quarks
using the clover action with 
Fermilab interpretation~\cite{Bailey:2012rr,Bazavov:2019aom}.
There was a $n_f=2$ determination by the ETMC 
collaboration~\cite{Atoui:2013zza} using twisted Wilson quarks, in which
they also determined the ratio of the tensor form factor to $f_+$ near
zero recoil. The RBC/UKQCD collaboration presented preliminary results
in~\cite{Flynn:2016vej,Flynn:2019any} and these were updated at this
conference~\cite{Flynn:2019jbg}. 
The preliminary results for their form factors 
are shown in Fig.~\ref{fig:BstoDs_rbc}.

\begin{figure}[t]
\centering
\includegraphics[width=0.75\textwidth]{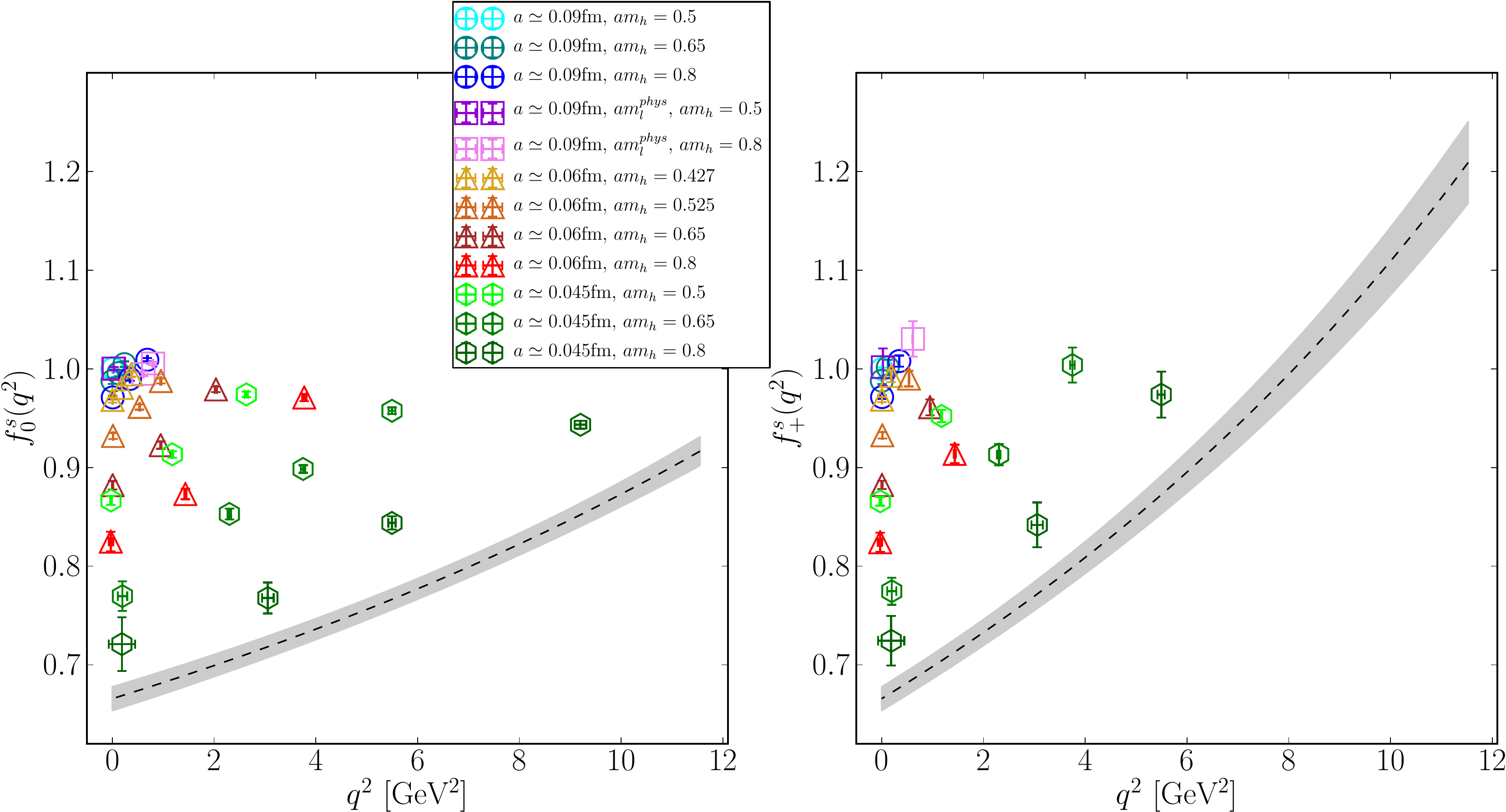}
\caption{Figure from \cite{McLean:2019qcx} showing results for
$B_s \to D_s$ $f_{0/+}$ form factors using the `heavy-HISQ' approach.
The colored open symbols show raw data for form factors calculated
at unphysically light $b$-quark masses on a range of ensembles with
lattice spacings from $a \sim 0.09$ -- 0.045 fm. The continuum extrapolated
results for the form factors at physical $b$-quark mass are given by the
gray bands.
\label{fig:BsDs_hpqcd-1}}
\end{figure}

\begin{figure}[t]
\centering
\includegraphics[width=0.43\textwidth]{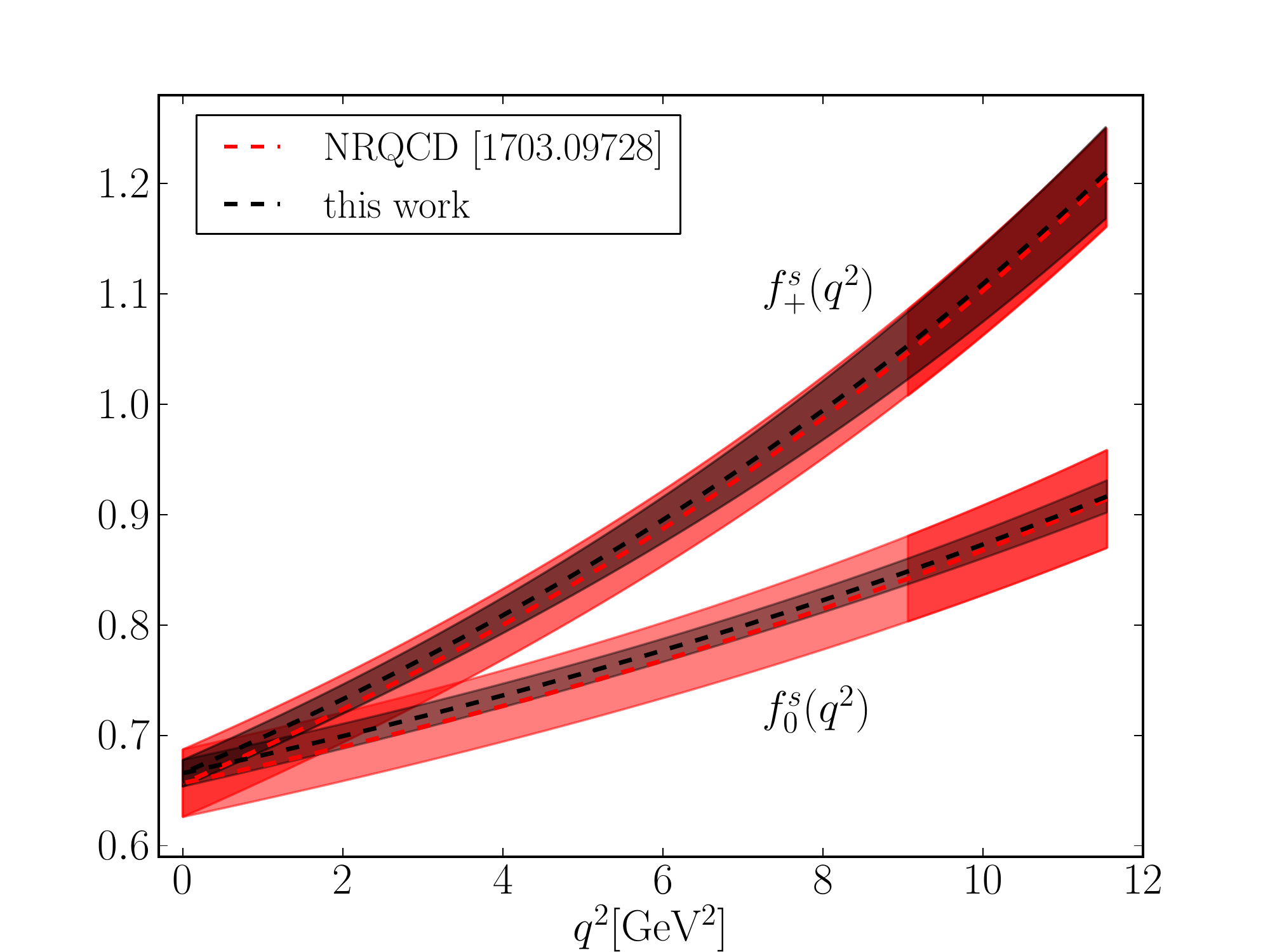} 
\includegraphics[width=0.43\textwidth]{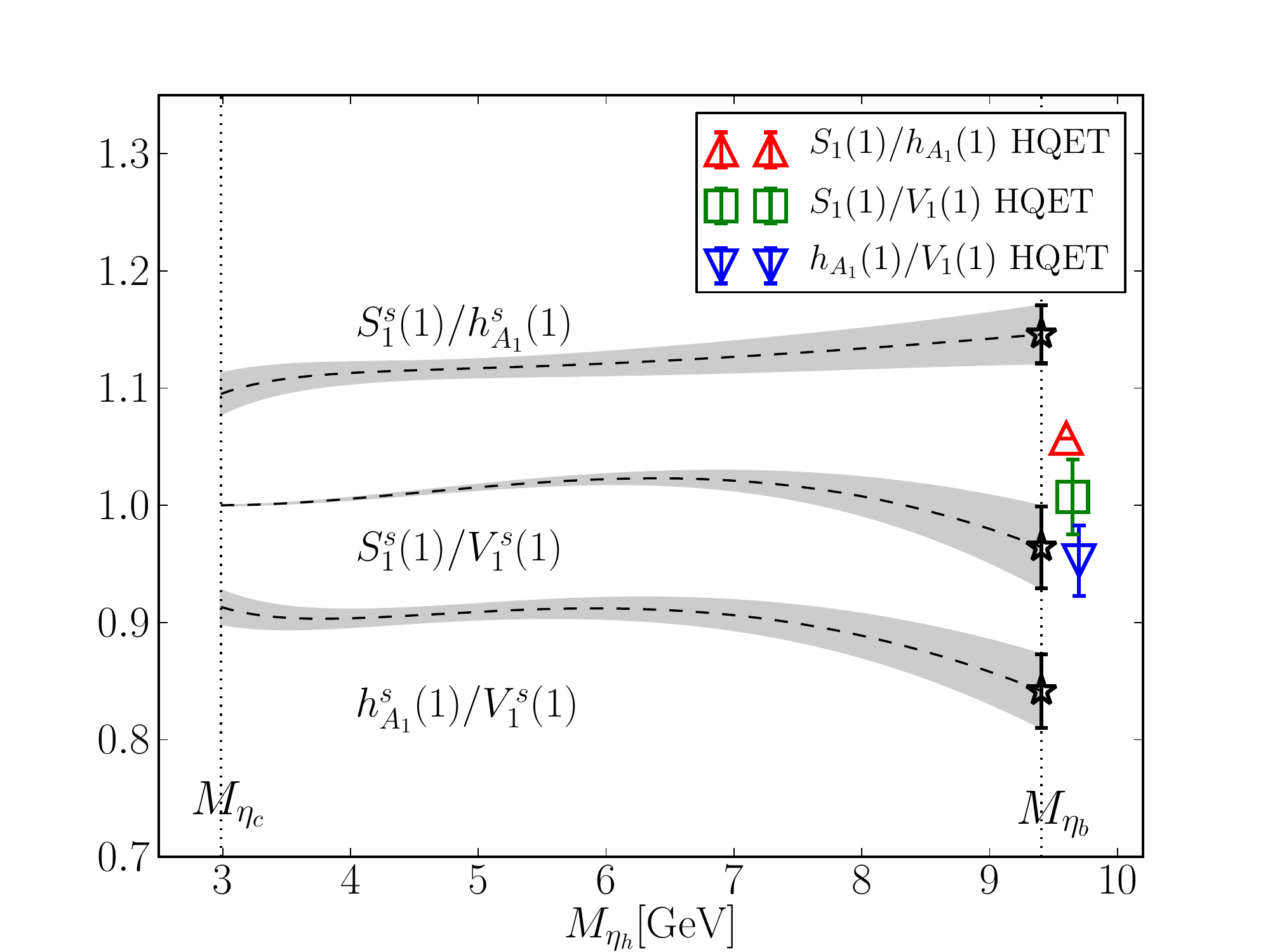} \\
\includegraphics[width=0.43\textwidth]{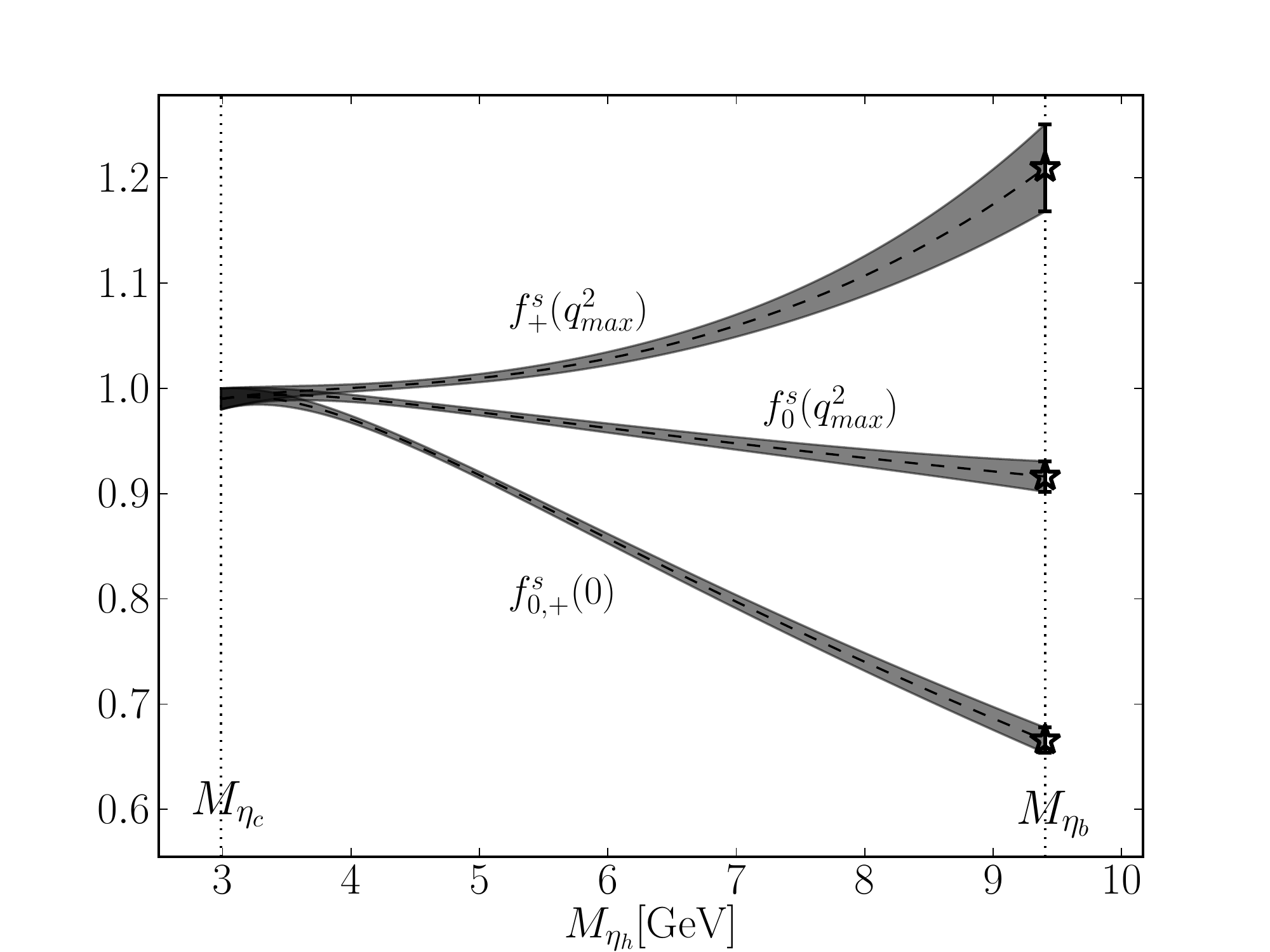}
\includegraphics[width=0.40\textwidth]{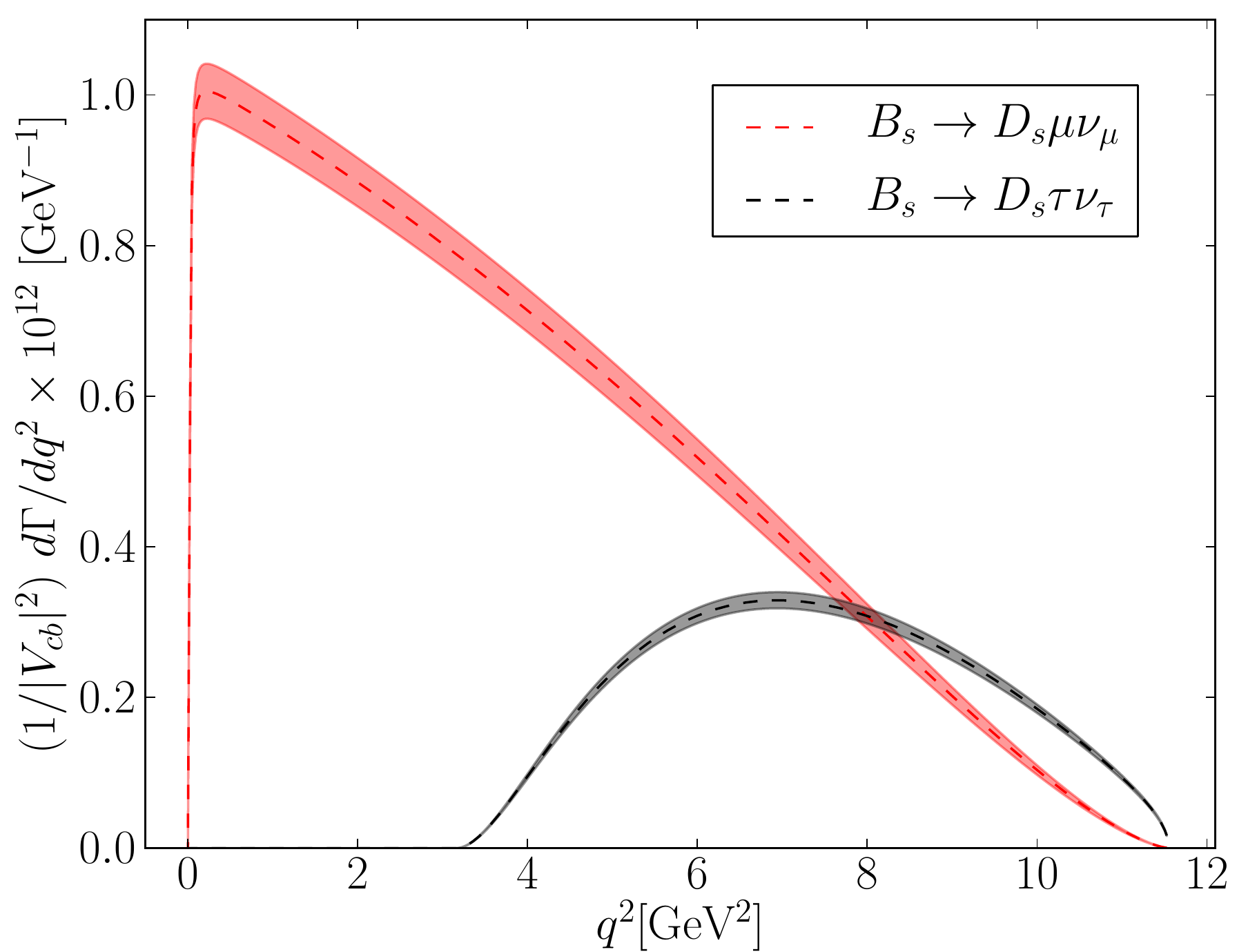}
\caption{
(Top-left) HPQCD's $B_s \to D_s$ form factors $f_{0/+}$ from~\cite{McLean:2019qcx}
using the `heavy-HISQ' approach compared with results from 
NRQCD~\cite{Monahan:2017uby}. (Top-right) Results for ratios of 
zero-recoil form factors as a function of $M_{\eta_h}$ used as a
physical proxy for the heavy quark mass $m_h$ used in simulation,
compared with predictions of HQET~\cite{Bernlochner:2017jka}.
(Bottom-left) Evolution of the form factor endpoints 
$f_{0/+}(0, q^2_{\text{max}})$ as a function of $M_{\eta_h}$.
(Bottom-right) Partial differential decays widths for 
$B_s \to D_s \, l \nu_l$ for $l=\mu,\tau$.
\label{fig:BsDs_hpqcd-2}}
\end{figure}

Recently the $f_{0/+}$ form factors
were computed over the entire kinematic range
using the heavy-HISQ approach by the 
HPQCD collaboration~\cite{McLean:2019qcx}. The raw data at unphysically
light $b$ mass and the form factors extrapolated to the $b$ mass are
shown in Fig.~\ref{fig:BsDs_hpqcd-1}. HPQCD also determined both form
factors using NRQCD $b$ in~\cite{Monahan:2017uby}; the results from both 
calculations are shown in Fig~\ref{fig:BsDs_hpqcd-2}.

Until recently the lattice QCD results for $B_s \to D_s^{(*)}$ form factors
could not be compared with experiment, that changed 
with the LHCb measurement~\cite{Aaij:2020hsi},
resulting in a new determination of 
$|V_{cb}|$ based on $B_s$ decays.
Their analysis was performed using both BGL and CLN parameterizations,
and the extracted $|\Vcb|$ is compatible between the two within errors.
Their result is compatible with both inclusive
and exclusive determinations from $B$ decays, but with larger errors.
These encouraging results increase the urgency for $B_s \to D_s^*$ results
away from zero recoil and increasing precision in both channels
$B_s \to D_s^{(*)}$.

\subsection{$B_c \to J/\psi$}
\begin{figure}
\centering
\includegraphics[width=0.45\textwidth]{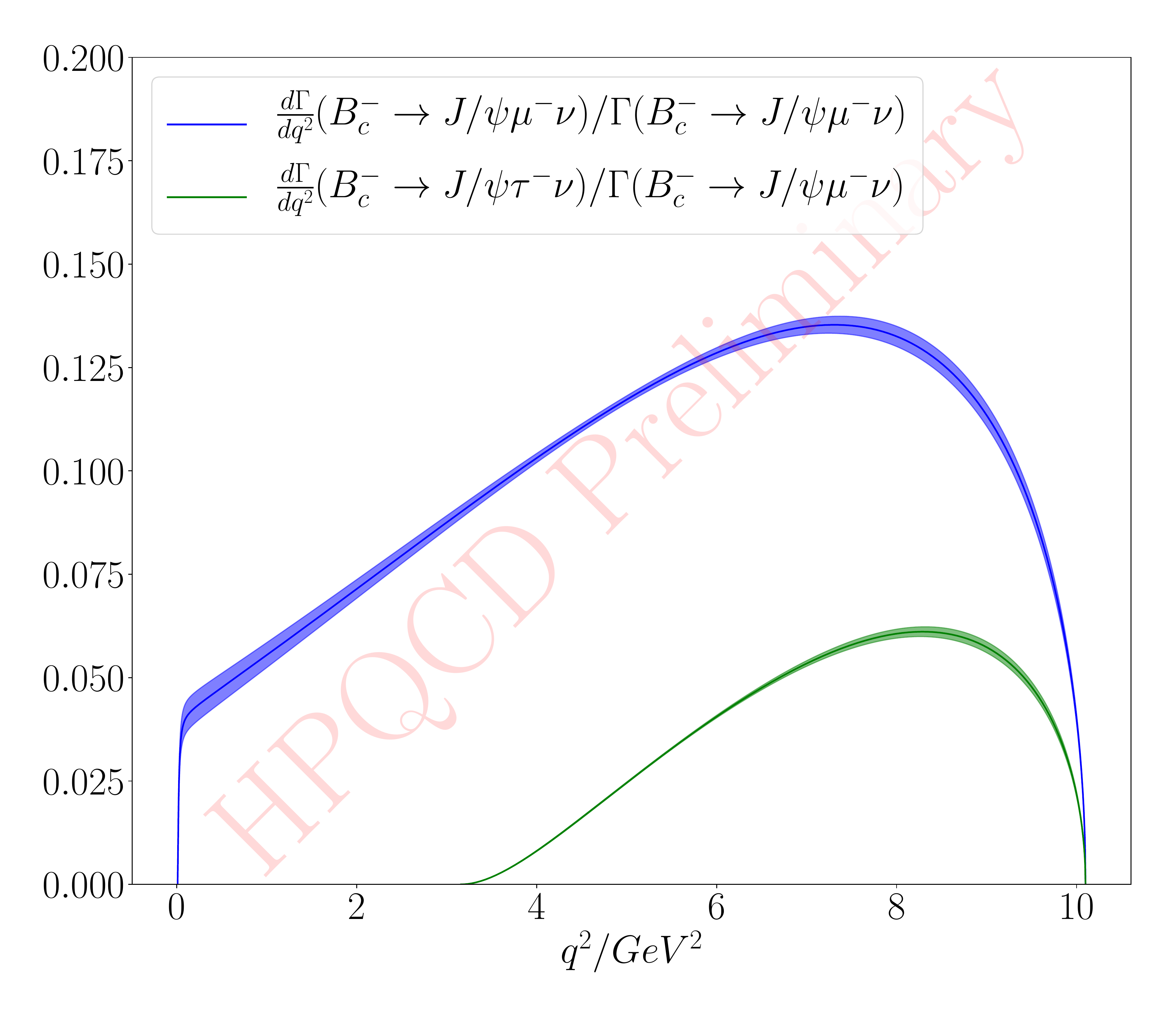}
\caption{
Differential decay widths for $B_c \to J/\psi \, l \nu$
with $l = \mu$ (blue) and $l = \tau$ (green), computed 
from lattice QCD using the `heavy-HISQ' methodology.
Figure courtesy J.~Harrison.
\label{fig:BcJpsi}}
\end{figure}

There are currently only preliminary results available for the 
$B_c \to J/\psi \, l \nu$ lattice form 
factors~\cite{Harrison:2019,Lytle:2016ixw,Colquhoun:2016osw}, by
HPQCD using the `heavy-HISQ' approach. 
The $R$-ratio for this decay was measured by LHCb~\cite{Aaij:2017tyk},
who found $R(J/\psi) = 0.71(17)_{\text{stat}}(18)_{\text{syst}}$. 
This value is $\sim 2 \sigma$ larger than what is expected in the SM
and although this value has large uncertainties it is
desirable to have a lattice determination,
particularly as the experimental precision improves 
(see Fig.~\ref{fig:RX_projection_stat_syst}). A preliminary
value $R(J/\psi) = 0.2592(92)$ was given in~\cite{Harrison:2019}, and
Fig.~\ref{fig:BcJpsi} shows the differential decay width as a function
of $q^2$.

\section{Conclusions} \label{Conclusions}
I would hazard that the study of $b \to c$ transitions is at somewhat of a
crossroads. There are several long-standing puzzles in this sector
where experimental data and theoretical predictions do not quite square,
and there are a number of welcome developments on the horizon that will be 
essential to a precise understanding that can either confirm or rule out
these discrepancies. Among these are improved predictions from the lattice
community over a larger kinematic range than has heretofore been available,
and results in new channels $B_s \to D_s^{*}$ and $B_c \to J/\psi$, and
also in the baryon sector, that can match experimental breakthroughs from LHC.
With the imminent results from Belle II,
$B \to D^{*}$ will surely remain the gold standard for extractions of 
$|V_{cb}|$,
and it is therefore a challenge for the lattice community to put these
calculations on a solid footing, and in particular away from zero recoil. 
As reviewed briefly above, fortunately
there are several lattice collaborations that have embarked upon this 
endeavour. 
\vskip -1cm
\section{Acknowledgements} \label{Acknowledgements}
I would like to thank the organizers of Lattice 2019 for an enjoyable
conference and the opportunity to present this material. 
I would like to thank C.~Davies, G.~de~Divitiis, B.~Dey, 
J.~Harrison, S.~Hashimoto, W.~Lee, L.~Lellouch, S.~Meinel, 
A.~Vaquero, and A.~Vladikas for useful discussions and insight.


\end{document}